\documentclass[12pt]{elsart}
\usepackage{graphicx}
\usepackage{amssymb}
\usepackage{amsmath}
\usepackage{hyperref}

\journal{Int. J. Theor. Phys.}

\newcommand{\be}{\begin{equation}}
\newcommand{\ee}{\end{equation}}

\DeclareMathOperator{\tr}{tr}

\begin{document}
%
\begin{frontmatter}
\title{Super universality of
the quantum Hall effect and the ``large $N$ picture" of the
$\vartheta$ angle}
\author{A.M.M.~Pruisken \corauthref{lc1}}
\corauth[lc1]{Fax: +31 20 525 57 78}
\ead{pruisken@science.uva.nl}
\address{Institute for Theoretical Physics, University of
Amsterdam, Valckenierstraat 65, 1018 XE Amsterdam, The
Netherlands}


\begin{abstract}
It is shown that the ``massless chiral edge excitations" are an
integral and universal aspect of the low energy dynamics of the
$\vartheta$ vacuum that has historically gone unnoticed. Within
the $SU(M+N)/S(U(M) \times U(N))$ non-linear sigma model we
introduce an effective theory of ``edge excitations" that
fundamentally explains the quantum Hall effect. In sharp contrast
to the common beliefs in the field our results indicate that this
macroscopic quantization phenomenon is, in fact, a {\em super
universal} strong coupling feature of the $\vartheta$ angle with
the replica limit $M=N=0$ only playing a role of secondary
importance. To demonstrate super universality we revisit the large
$N$ expansion of the $CP^{N-1}$ model. We obtain, for the first
time, explicit scaling results for the quantum Hall effect
including quantum criticality of the quantum Hall plateau
transition. Consequently a scaling diagram is obtained describing
the cross-over between the weak coupling ``instanton phase" and
the strong coupling ``quantum Hall phase" of the large $N$ theory.
Our results are in accordance with the ``instanton picture" of the
$\vartheta$ angle but fundamentally invalidate all the ideas,
expectations and conjectures that are based on the historical
``large $N$ picture."
\end{abstract}


\end{frontmatter}
\clearpage\tableofcontents \clearpage
\vspace{2cm}
\section{Introduction}
\subsection{Super universality}
In a series of investigations on the grassmannian $SU(M+N)/S(U(M)
\times U(N))$ non-linear sigma model in two dimensions it was
shown that the $\vartheta$ vacuum generally displays massless
excitations that propagate along the confining ``edge" of the
system.\cite{PBSI,PBSIII,SP,PruiskenShankarSurendran,PruiskenShankarSurendran-1}
This new aspect of the sigma model has come in many ways as a
welcome surprise. In applications to quantum spin liquids,
\cite{Haldane} for example, the edge excitations describe the
dynamics of the ``dangling" quantum spins located at the ``edges"
of the spin chain.
\cite{PruiskenShankarSurendran,PruiskenShankarSurendran-1} In the
context of quantum Hall liquids, \cite{Pruisken84} on the other
hand, these excitations are identically the same as those
described by the theory of ``chiral edge bosons." \cite{PBSIII,SP}
Quite similar to the semiclassical ideas that are popularly used
for quantum Hall systems \cite{Prange-Prange} one may formulate a
percolating network of ``edge" excitations while ignoring all the
other excitations in the problem. One can then show that in the
limit of large distances the network model is identically the same
as the original non linear sigma model. \cite{PBSIII,SP} The
``edge" of the $\vartheta$ vacuum is therefore the key for
resolving longstanding issues such as the cross-over between {\em
percolation} and {\em localization} which is known to complicate
the experiment on scaling conducted on realistic quantum Hall
samples. \cite{Schaijk}

Massless chiral edge excitations furthermore lay the bridge
between the $\vartheta$ angle concept on the one hand, and the
phenomenological approaches to the fractional quantum Hall effect
based on Chern-Simons gauge theory on the other.\cite{Wen} This
has led, amongst many other things, to a complete Luttinger liquid
theory of edge excitations that includes the effects of disorder,
the Coulomb interaction as well as the coupling of the theory to
external potentials. \cite{SP}

These specific examples clearly indicate that the topological
concept of a $\vartheta$ angle is much richer and more profound
than previously thought. Notice that the physics of the ``edge"
would already be a useful advance even if its relevance was
limited to the two dimensional electron gas or quantum spin chains
alone. This kind of knowledge becomes only more interesting,
however, if it turns out that the quantum Hall effect has, in
fact, a much more general significance that eventually could shed
some new light on the strong coupling problems in QCD where the
topological concept of a $\vartheta$ vacuum arose first.
\cite{Rajaraman} Within the grassmannian sigma model one finds,
for example, that the $SU(M+N)$ symmetry is spontaneously broken
at the ``edge" of the system whereas the critical correlations
along the ``edge" are identically the same for all values of $M$
and $N$. \cite{PBSIII} These unexpected features have motivated
several studies where the idea of {\em super universality} has
emerged.
\cite{PruiskenShankarSurendran,PruiskenShankarSurendran-1,PruiskenBurmistrov,PruiskenBurmistrov1,PruiskenBurmistrov2}
The essence of this idea is that the $\vartheta$ vacuum displays
all the basic aspects of the quantum Hall effect for all
non-negative values of $M$ and $N$. These include not only the
massless chiral edge excitations but also the existence of {\em
robust} topological quantum numbers that explain the precision and
stability of the quantum Hall plateaus, as well as the existence
of {\em gapless bulk excitations} at $\vartheta=\pi$ that describe
a quantum phase transition between adjacent plateaus.

The statement of super universality is important because it
explains in a natural manner why completely different theories of
the $\vartheta$ angle display the same physical phenomena. It
encompasses the concept of ordinary universality in critical
phenomena phenomenology which is a statement made on critical
exponent values alone. Quantum criticality at $\vartheta=\pi$ may
in principle be different depending on the specific application of
the $\vartheta$ angle that one is interested in. This aspect of
the problem is in many ways the same as quantum criticality in
$2+\epsilon$ dimensions where each value of $M$ and $N$ is known
to describe a different universality class.\cite{Brezin}

Several interesting examples of super universality have already
emerged in recent years. We mention, in particular, the
Finkelstein approach to localization and interaction effects
\cite{PruiskenBurmistrov1} which explains why the infinitely
ranged Coulomb interaction does not affect the basic phenomena of
scaling as predicted by the free electron gas \cite{Pruisken88}
and observed in the experiment.\cite{Wei} The Finkelstein
approach, however, fundamentally alters our understanding of the
quantum critical behavior of the electron gas. This behavior
actually belongs to a novel non-Fermi liquid universality class
with a different meaning for the critical exponents
\cite{PruiskenBurmistrov1,PBSII} and characterized by a previously
unrecognized interaction symmetry termed $\mathcal{F}$ invariance.
\cite{PBSI} A second example is the Ambegaokar-Eckern-Sch\"{o}n
model of the Coulomb blockade.\cite{AES} This theory of the
$\vartheta$ angle is perhaps the simplest of all since it involves
a single abelian field variable in one dimension. Yet it shows all
the richly complex physics of quantum Hall liquids and quantum
spin liquids.\cite{PruiskenBurmistrov2}
\subsection{Large $N$ expansion}
In this investigation we embark on a third example of super
universality, the large $N$ expansion of the $CP^{N-1}$ model
which is obtained from the grassmannian theory by putting $M,N$
equal to $1,N-1$. This specific case is interesting because it is
one of view places in the theory where the $\vartheta$ vacuum is
accessible from the strong coupling side. Even though the matter
has been studied in detail and elaborated upon a long time ago,
\cite{Dadda,Witten,Affleck80} it turns out that the physics of the
``edge" is a source of unforseen and troublesome complexity that
has direct consequences for our understanding of the theory as a
whole. We will show that the large $N$ steepest descend
methodology, which is standardly performed for an infinite system,
misses all the subtle aforementioned features of the confining
``edge" of the $\vartheta$ vacuum. The historical papers on the
subject mishandle the ``massless chiral edge excitations" in the
problem and, hence, the most interesting aspect of the $\vartheta$
angle, the quantum Hall effect, remained concealed. The large $N$
analysis that follows is in many ways a completely novel theory.
It not only resolves many longstanding controversies on
topological issues in both quantum field theory \cite{Coleman1}
and condensed matter theory
\cite{Affleck83,Affleck91,Verbaarschot,Weidenmueller,Zirnbauer}
but also demonstrates in an unequivocal manner that the
$\vartheta$ angle is, in fact, the fundamental theory of the
quantum Hall effect.\cite{Pruisken84,Prange}

To obtain the correct low energy dynamics of the $\vartheta$
angle, notably the quantum Hall effect, one must handle the large
$N$ steepest descend methodology very differently from what has
been done before. Since the massless ``edge" excitations are
distinctly different from those of the ``bulk" of the system they
should be disentwined and studied separately. This can in general
been done because of their simple topological properties.
\cite{PBSIII} More specifically, the ``edge" excitations generally
have a {\em fractional} topological charge whereas the ``bulk"
excitations always carry a strictly {\em integral} topological
charge. Separating the ``edge" from the ``bulk" is therefore
synonymous for separating the {\em fractional} topological sectors
of the theory from the {\em integral} sectors.

The crux of this investigation is the introduction of an effective
theory of ``edge" excitations that is obtained by formally
eliminating the ``bulk" degrees of freedom. This effective theory
expresses the low energy dynamics of the $\vartheta$ vacuum in
terms of two ``physical observables" only. These two quantities
have previously been identified as the longitudinal conductance
and Hall conductance respectively in the context of the disordered
electron gas. \cite{Pruisken84,Prange} We present a comprehensive
study of these physical quantities on both the weak and the strong
coupling side of the problem. The picture that emerges is
precisely in accordance with the renormalization group ideas on
the quantum Hall effect that have originally been proposed on the
basis of the semiclassical instanton methodology
alone.\cite{Pruisken88,Pruisken85,Pruisken87} Unlike the previous
situation, however, we now have - for the first time - an explicit
demonstration of the robust quantization of the Hall conductance
together with explicit scaling results for the quantum Hall
plateau transition.
\subsection{Outline of this work}
In order to properly account for the new physics associated with
the ``edge," we will present our findings for the large $N$
expansion in a step by step manner. We start out, in Section
\ref{Sec2}, with a brief summary of the microscopic origins of the
$\vartheta$ angle followed by a brief introduction to subject of
massless chiral edge excitations. In Section \ref{Sec3} we embark
on the general question of how to disentwine the massless edge
excitations from the bulk degrees of freedom. The effective theory
of massless ``edge" excitations is introduced in Section
\ref{finite-size}. This effective theory leads directly to a
generalized Thouless criterion for the quantum Hall effect which
relates the generation of a mass gap for bulk excitations to the
insensitivity of the system to changes in the boundary conditions.
The argument is based on very general principles only and
therefore sets the stage for the concept of super universality.

After these preliminaries we specialize to the large $N$ expansion
of the $CP^{N-1}$ model. In Section \ref{Observables} we elaborate
on the results recently obtained from the instanton calculational
technique \cite{PruiskenBurmistrov} which is the starting point of
the remainder of this paper. In Section \ref{Sec4} we review the
standard large $N$ saddle point methodology and show that it
conflicts with super universality. We then point out, in Section
\ref{Quantum-Hall}, that the idea of the massless ``edge"
excitations fundamentally alters the structure of the steepest
descend methodology with direct consequences for the ``bulk" of
the system. We evaluate the effective theory of massless chiral
edge excitations and obtain explicit scaling results for the
quantum Hall effect and the quantum Hall plateau transitions that
were previously invisible. In Section \ref{cond-distr} we
elaborate on the physics of the plateau transitions and show that
they are a prototypical example of broad ``conductance"
distributions in the quantum theory of metals. In Section
\ref{twist} we show how the effective theory of the ``edge" can be
used as a important check on the Levine-Libby-Pruisken argument
for {\em de}-localized or gapless ``bulk" excitations at
$\vartheta=\pi$. \cite{Pruisken84} These excitations do exist in
the large $N$ theory even though the transition is a first order
one.

In Section \ref{Pseudo-instanton} we embark on the cross-over
between the weak and strong coupling phases of the large $N$
theory. We first show that the results of the large $N$ steepest
descend methodology can in general be decomposed in a discrete set
of topological sectors in complete accordance with the
semiclassical theory based on instantons.
\cite{Rajaraman,BergLuesher,Jevicki} We then show that the
``dilute instanton gas" expressions for the free energy and the
renormalization group $\beta$ functions retain their general form
in the entire range from weak coupling all the way down to the
strong coupling phase of the large $N$ theory. The main results of
this paper are summarized by the renormalization group flow
diagram of the ``conductances" plotted in Fig. \ref{FIG4}. This
paper ends (Section \ref{Summary}) with a summary of the results
and a conclusion.
\section{The $\vartheta$ angle and physics of the ``edge" \label{Sec2}}
The non-linear sigma model representation of Anderson localization
in two dimensions and in a perpendicular magnetic field is
discussed in detail in Ref. \cite{Prange}. It involves the
grassmannian field variable $Q$ with $Q^2 = {\bf 1}_{M+N}$ that
can be written in a standard fashion as follows
\be
 Q= T^{-1} \Lambda T .\label{Q}
\ee
Here, $T \in SU(M + N )$ and $\Lambda$ denotes a diagonal matrix
with $M$ elements $+1$ and $N$ elements $-1$
\begin{equation}
\Lambda = \begin{pmatrix}
  \mathbf{1}_{M} & 0 \\
  0 & -\mathbf{1}_{N}
\end{pmatrix}. \label{Lambda}
\end{equation}
The action of the electron gas reads
\begin{eqnarray}
 S [Q] &=& S_\sigma [Q] + \pi \omega \rho \int d^2 x \, \tr\Lambda Q \label{SeffStart-0}\\
 S_\sigma [Q]& = & - \frac{1}{8} g_0 \int d^2 x \,\tr \partial_\mu Q \partial_\mu Q
 + \frac{1}{8}\sigma^0_{H} \int d^2 x \,\tr \varepsilon_{\mu\nu}
 Q\partial_\mu Q\partial_\nu Q .\label{SeffStart}
\end{eqnarray}
The dimensionless quantities $g_0$ and $\sigma_{H}^0$ are the mean
field parameters for {\em longitudinal} conductance and {\em Hall}
conductance respectively in units of $e^2/h$. The quantity $\rho$
denotes the density of electronic levels and $\omega$ the external
frequency.

The success of the non-linear sigma model representation of
Anderson localization ultimately relies on our ability to evaluate
the Kubo expressions for the macroscopic conductances. These
quantities are usually defined for a fixed frequency $\omega$ and
an arbitrarily large sample size. They are most elegantly
represented in terms of a background matrix field $t (x) \in
SU(M+N)$ that varies slowly in space. The master formulae for the
background field action reads as follows
\begin{eqnarray}
 \exp \{-\mathcal{F} + S^\prime_\sigma [t] \} &=& \int \mathcal{D} [Q] \exp \left\{
 S_\sigma [t^{-1} Q t] + \pi \omega \rho \int d^2 x \, \tr\Lambda Q
 \right\}.\label{master}
\end{eqnarray}
Here, $\mathcal{F}$ denotes the free energy and the action
$S^\prime [t]$ is the shift away from the equilibrium distribution
as a result of the background field insertion $t (x)$. Provided $t
(x)$ satisfies the classical equations of motion this action takes
on the form of the sigma model itself
\begin{eqnarray}
 S^\prime_\sigma [t] &=& - \frac{1}{8} g^\prime (\omega) \int d^2 x \,
 \tr \partial_\mu v \partial_\mu v + \frac{1}{8}\sigma^\prime_{H} (\omega) \int d^2 x \,
 \tr \varepsilon_{\mu\nu} v \partial_\mu v \partial_\nu v \label{S-prime}
\end{eqnarray}
with ${v} = t^{-1} \Lambda t$. This result is the only possible
local action with at most two derivatives that is compatible with
the symmetries of the problem. One therefore expects that Eq.
\eqref{S-prime} has a quite general significance that is
independent of $M$ and $N$. In particular, the quantities of
physical interest are $g^\prime (\omega)$ and $\sigma^\prime_{H}
(\omega)$ in Eq. \eqref{S-prime} which in the replica limit $M,N
\rightarrow 0$ precisely correspond to the Kubo expressions for
linear response averaged over the impurity ensemble.
\subsection{Spontaneous symmetry breaking at the ``edge" \label{edge-spin}}
The massless excitations along the ``edge" of the $\vartheta$
vacuum were recognized first in Refs \cite{PBSIII,SP}. To
understand this important aspect of the problem we consider the
simplest possible scenario of an electron gas in a strong
perpendicular magnetic field such that the disordered Landau bands
are well separated. By taking the Fermi level $E_F$ inside a
Landau gap then both quantities $g_0$ and $\rho$ in Eq.
\eqref{SeffStart} become zero as it should be. The mean field Hall
conductance, however, is strictly integer valued $\sigma_{H}^0 =k$
with $k$ denoting the number of completely filled Landau levels.
\cite{Prange} The action of Eq. \eqref{SeffStart} now arises
solely from the one dimensional ``edge" of the system and we can
write
\begin{eqnarray}
 S [Q] \rightarrow S_{gap} [Q] &=& \frac{k}{2} \oint d x\, \tr T \partial_x T^{-1}
 \Lambda  + \pi\omega\rho_\textrm{edge}
 \oint d x\, \tr\Lambda Q .\label{SedgeIsoA}
\end{eqnarray}
Here, we have made use of the fact that the topological charge can
be written as an integral along the edge $\oint$ according to
\begin{eqnarray}
 \mathcal{C} [Q] = \frac{1}{16\pi i} \int d^2 x \tr
 \epsilon_{\mu\nu} Q \partial_\mu Q \partial_{\nu} Q
 &=& \frac{1}{4\pi i} \oint d x\, \tr T \partial_x T^{-1}
 \Lambda .\label{top-charge}
\end{eqnarray}
The extra term with $\rho_\textrm{edge}$ in Eq. \eqref{SedgeIsoA}
indicates that although there are no electronic levels near $E_F$
in the bulk of the system there is nevertheless a finite density
of ``edge states" that carry the Hall current.  Surprisingly, this
one dimensional theory is critical and exactly
solvable.\cite{PBSIII} It can be shown that Eq. \eqref{SedgeIsoA}
is completely equivalent to the theory of chiral edge bosons with
the drift velocity $v_d$ of the chiral edge electrons given by $
v_d  = {k}/{ 2\pi \rho_\textrm{edge}}$. \cite{SP} Some important
correlations are as follows \cite{PBSIII}
\be
 \langle Q \rangle_\textrm{edge} = \Lambda
 \label{edge-correl-1}
\ee
indicating that the $SU(M+N)$ symmetry is spontaneously broken at
the edge of the $\vartheta$ vacuum. Furthermore
\be
 \langle Q^{+-}_{\alpha\beta} (x) Q^{-+}_{\beta\alpha} (x^\prime)
 \rangle_\textrm{edge} =4\vartheta(x^\prime-x)
 e^{-\omega (x^\prime-x)/v_d} \label{edge-correl-2}
\ee
where $Q^{+-}_{\alpha\beta}$ and $Q^{-+}_{\alpha\beta}$ denote the
$M\times N$ components in the off-diagonal blocks and
$\vartheta(x)$ is the Heaviside step function. The remarkable and
surprising feature of these critical edge correlations is that
they are completely independent of $M$ and $N$. Moreover, if we
interpret the edge coordinate $x$ as the imaginary time $\tau$
then one recognizes Eq. \eqref{SedgeIsoA} as the bosonic path
integral of an $SU(M+N)$ spin with quantum number $s=k/2$ in a
magnetic field $B=\rho_{edge} \omega$.
\cite{PruiskenShankarSurendran}

Next, to compute the conductances we go back to our master
formulae of Eqs \eqref{master} and \eqref{S-prime} and write
\begin{eqnarray}
 \exp \{-\mathcal{F} + S^\prime_\sigma [t] \} &=& \int \mathcal{D} [Q] \exp \left\{
 S_\sigma [t^{-1} Q t] + \pi \omega \rho \oint d x \, \tr\Lambda Q
 \right\} \nonumber \\
 &=& \int \mathcal{D} [Q] \exp \left\{
 S_{gap}  [ Q ] + \frac{k}{2} \oint d x \, \tr t \partial_x t^{-1} Q
 \right\} .\label{master-edge}
\end{eqnarray}
We immediately obtain
\be
 S^\prime_\sigma [t] = \frac{k}{2} \oint d x \, \tr t \partial_x t^{-1} \langle
 Q \rangle_{edge} = 2\pi i k \mathcal{C} [v] \label{S-prime-edge-1}
\ee
where we have used Eqs \eqref{top-charge} and
\eqref{edge-correl-1}. The conductance parameters are therefore
independent of $M$ and $N$ and given by
\be
 g^\prime (\omega) = 0 ~~,~~~ \sigma_H^\prime (\omega) = k .
 \label{qHe-edge-1}
\ee
We see that the quantum Hall effect reveals itself through
spontaneous symmetry breaking at the ``edge" of the $\vartheta$
vacuum. This unexpected feature of the sigma model in two
dimensions has historically been overlooked. From now onward we
recognize the action of Eq. \eqref{SedgeIsoA} as the critical
fixed point action of the quantum Hall state.
\section{Disentangling the bulk and the edge \label{Sec3}}
\subsection{The matrix field variable $Q$}
To disentwine the massless ``edge" excitations from the ``bulk"
excitations we write
\be
 Q = \varphi^{-1} Q_0 \varphi \label{tQ0t}
\ee
where $Q_0 = T_0^{-1} \Lambda T_0$ has a fixed value $Q_0
=\Lambda$ at the edge. The matrix $\varphi \in SU(M+N)$ generally
stands for the ``fluctuations" about these special boundary
conditions. These distinctly different components of $Q$ are
termed the ``bulk" component and ``edge" component respectively.
They are topologically classified according to
\be
 \mathcal{C} [Q] = \mathcal{C} [Q_0] + \mathcal{C} [ q ] ~~, ~~~
 q=\varphi^{-1} \Lambda \varphi
 \label{Q0+t}
\ee
where $\mathcal{C} [Q_0]$ is by construction an {\em integer} and
$-\frac{1}{2} < \mathcal{C} [q] \leq \frac{1}{2}$ stands for the
{\em fractional} part of $\mathcal{C} [Q]$. Notice that for the
special case of integer filling fractions $\nu$ the action (Eq.
\ref{SedgeIsoA}) solely depends on the ``edge" component $\varphi$
of $Q$
\begin{eqnarray}
 S_{gap} [Q] = S_{gap} [\varphi^{-1} Q_0 \varphi] & = & \frac{k}{2}
 \oint d x\, \tr \left[ \varphi
 \partial_x \varphi^{-1}  + \frac{\omega}{v_d} q \right]\Lambda
 + 2\pi i k \mathcal{C} [Q_0] .\label{S-gap}
\end{eqnarray}
The ``bulk" component $Q_0$ gives rise to a trivial phase factor
$2\pi i k \mathcal{C} [Q_0]$ and can be ignored.
\subsection{The $\vartheta$ parameter or $\sigma_H^0$}
\begin{figure}
\includegraphics[width=80mm]{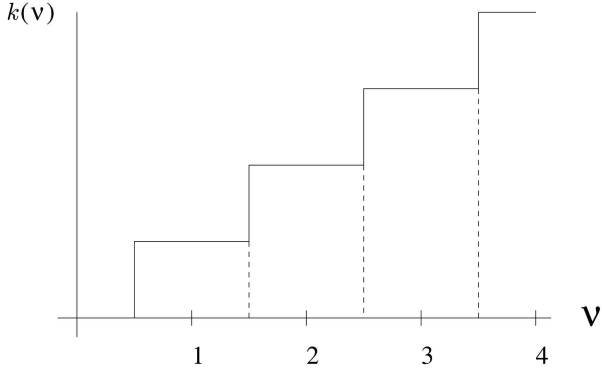}
\caption{ The edge part of $\sigma_{H}^0$ with varying $\nu$, see
text.} \label{FIG1}
\end{figure}

\begin{figure}
\includegraphics[width=80mm]{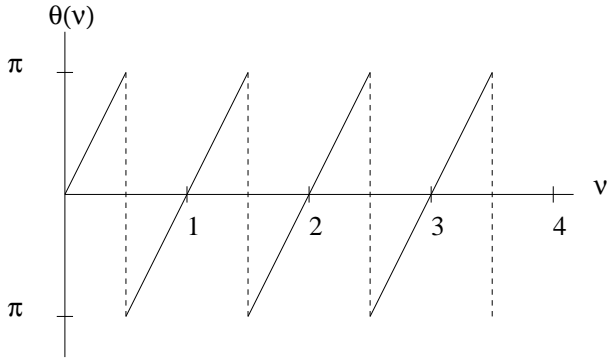}
\caption{ The bulk part of $\sigma_H^0$ with varying $\nu$, see
text. } \label{FIG2}
\end{figure}

Next, to address the general theory of Eq. \eqref{SeffStart} it is
convenient split the mean field parameter $\sigma_H^0 =\nu$ into
an {\em integral} piece $k(\nu)$ and a {\em fractional} piece
$-\pi < \theta (\nu) \leqq \pi$ according to (see Figs \ref{FIG1}
and \ref{FIG2})
\be
 \sigma_{H}^0 = \nu = k(\nu) + \frac{\theta(\nu)}{2\pi}.
 \label{split3}
\ee
Microscopically, i.e. for the electron gas in strong magnetic
fields, the quantities $k(\nu)$ and $\theta(\nu)$ appear as
distinctly different contributions from the ``edge" and the
``bulk" respectively. \cite{Pruisken88-1} The mean field
parameters $g_0 = g_0 (\theta(\nu))$ and $\rho =
\rho(\theta(\nu))$ are typical ``bulk" quantities that for a given
Landau band depend on $\theta(\nu)$ only. They are symmetric under
$\theta(\nu) \Leftrightarrow -\theta(\nu)$ which is termed
``particle-hole" symmetry.

The split in Eq. \eqref{split3} can easily be understood by
considering a system with a ``clean" confining edge that is
spatially separated from the homogenously distributed disorder in
the interior of the system. This kind of thought experiment has
frequently been used in Refs~\cite{PBSIII,SP} and here it shows
that the integral piece $k(\nu)$ typically arises from the ``edge"
states in the problem that are unaffected by the disorder whereas
$\theta (\nu)$, like $g_0 (\theta (\nu))$ and $\rho (\theta
(\nu))$, is entirely determined by the disorder in the ``bulk" of
the system.
\subsection{Actions for the ``bulk" and the ``edge"}
Using these definitions we next split Eqs \eqref{SeffStart-0} and
\eqref{SeffStart} into a ``bulk" part $S_{bulk}$ and an ``edge"
part $S_{edge}$ as follows
\begin{eqnarray}
 S [Q] &=& S_{bulk} [Q] + S_{edge} [q] \label{bulk+edge} \\
 S_{bulk} [Q] &=& \tilde{S}_\sigma [Q] +  \pi\omega\rho \int d^2 x\, \tr Q \Lambda
 \label{Bulk-1} \\
 S_{edge} [q] &=& 2\pi i k(\nu) \mathcal{C} [q] +  \pi\omega\rho_{edge} \oint d x\, \tr q
 \Lambda .\label{Edge-1}
\end{eqnarray}

Here, $\tilde{S}_\sigma [Q]$ denotes the sigma model for the
``bulk" of the system
\begin{eqnarray}
 \tilde{S}_\sigma [Q] &=&  - \frac{g_0}{8} \int d^2 x\, \tr
 \partial_\mu Q \partial_\mu Q
 + i \theta(\nu) \mathcal{C} [Q] .~~~\label{Bulk-2}
\end{eqnarray}
By separating the integrals over the ``bulk" modes $Q_0$ and
``edge" components $q$ or $\varphi$ we can write
\begin{eqnarray}
 Z &=& \int \mathcal{D} [q] ~e^{S_{edge} [q]} ~ \int_{\partial V} \mathcal{D} [Q_0]
 ~e^{S_{bulk} [\varphi^{-1} Q_0 ~\varphi]} .
 \label{Z-1}
\end{eqnarray}
%
%
%

Here, the subscript $\partial V$ reminds us of the fact that the
functional integral over $Q_0$ has to be performed with fixed
boundary conditions $Q_0 = \Lambda$. Notice that for integer
filling fractions $\nu$ the action $S_{bulk}$ is zero and Eq.
\eqref{Z-1} stands for the critical theory of the ``edge" as it
should be. On the other hand, in the absence of the ``edge"
component $\varphi$ or $q$ we obtain a theory of pure ``bulk"
excitations $S_{bulk} [Q_0]$ with a strictly quantized topological
charge $\mathcal{C} [Q_0]$. This theory is precisely in accordance
with the semiclassical ``instanton picture" of the $\vartheta$
angle.

As a final remark, it should be mentioned that the results of this
Section have a quite general significance that is independent of
condensed matter applications. For example, the split written in
Eq. \eqref{bulk+edge} can always be made even though the
decomposition of Eq. \eqref{split3} may not always be physically
obvious. The main advantage of the electron gas, therefore, is
that the distinction between the ``bulk" and ``edge" naturally
emerges from the microscopic origins of the $\vartheta$ angle.
\subsection{Finite size scaling \label{finite-size}}
We are now in a position to introduce finite size scaling ideas
for the macroscopic conductances. For this purpose we take the
limit $\omega = 0$ in the action for the ``bulk" $S_{bulk}$ and
let the infrared be defined by the sample size $\lambda^\prime$.
Eq. \eqref{Z-1} can now be written in terms of an effective theory
for the ``edge" according to
\begin{eqnarray}
 Z &=& e^{-\mathcal{F}_b }\int \mathcal{D} [q]
 ~e^{S_{edge} [q]+ \tilde{S}^\prime_\sigma [q]} \label{bulk-Z-1}
\end{eqnarray}
where
\begin{eqnarray}
 e^{-\mathcal{F}_b + \tilde{S}^\prime_\sigma [q]}&=& \int_{\partial V} \mathcal{D} [Q_0]
 e^{\tilde{S}_{\sigma} [ \varphi^{-1} Q_0 ~ \varphi ]} .\label{bulk-Z-2}
\end{eqnarray}
Here, $\mathcal{F}_b = \mathcal{F}_b (\theta(\nu))$ denotes the
``bulk" free energy. Notice that the effective action
$\tilde{S}^\prime_\sigma [q]$ is, in effect, a measure for the
sensitivity of the ``bulk" of the system to infinitesimal changes
in the boundary conditions. Emerging from Eqs \eqref{bulk-Z-1} and
\eqref{bulk-Z-2} is therefore the well known ``scaling picture" of
Anderson localization where one divides the macroscopic system
into an array of much smaller ``blocks" of size $\lambda \ll
\lambda^\prime$. Provided the $\varphi$ field obeys the classical
equations of motion one can generally express the effective action
$\tilde{S}^\prime_\sigma [q]$ as follows
\begin{eqnarray}
 \tilde{S}^\prime_\sigma [q] &=&  - \frac{1}{8} g^\prime (\lambda) \int d^2 x\, \tr
 \partial_\mu q \partial_\mu q
 + i \theta^\prime (\lambda) \mathcal{C} [q] ~~~\label{Bulk-prime-2}
\end{eqnarray}
where $g^\prime (\lambda)$ and $\theta^\prime (\lambda)$ are the
``response" parameters associated with a single block. They are
explicitly given as correlations of the Noether current $J_\mu =
Q_0 \partial_\mu Q_0$ according to \cite{Pruisken87,Prange}
\begin{eqnarray}
 g^\prime (\lambda) &=& g_0 + \frac{ (g_0
 )^2}{8M N \lambda^2} \int_{x,x^\prime} \langle \tr J_\mu (x)
 J_\mu (x^\prime)\rangle \label{Kubo1} \\
 \theta^\prime (\lambda) &=& \theta(\nu) +
 \frac{ (g_0)^2}{8M N \lambda^2} \int_{x,x^\prime}
 \langle \tr \varepsilon_{\mu\nu} J_\mu (x) J_\nu (x^\prime)
 \Lambda \rangle\label{Kubo2}
\end{eqnarray}
where the expectations $\langle\dots \rangle$ are with respect to
$\tilde{S}_\sigma [Q_0]$ which is defined for a block of size
$\lambda$. Here, $g^\prime (\lambda)$ stands for the parallel
conductance and
\begin{eqnarray}
 \sigma^\prime_H (\lambda) &=& k(\nu) + \frac{\theta^\prime (\lambda)}{2\pi}
 ~~~\label{Hall-prime-1}
\end{eqnarray}
denotes the Hall conductance. By considering a sequence of scale
sizes $\lambda$, $2\lambda$, $4\lambda$ etc. then the results of
Eqs \eqref{Kubo1} - \eqref{Hall-prime-1} essentially tell us how
single blocks are being joined together to form the transport
parameters of bigger blocks. This scaling scenario immediately
suggests a {\em generalized Thouless criterion} for Anderson
localization in the presence of a magnetic field. More
specifically, since exponentially localized electronic levels are
insensitive to changes in the boundary conditions one expects that
the response parameters $g^\prime (\lambda)$ and $\theta^\prime
(\lambda)$ render exponentially small
\begin{eqnarray}
 g^\prime (\lambda) ,~ \theta^\prime (\lambda) \propto~\exp \{-\lambda/\xi
 \} \label{qHe-corrections}
\end{eqnarray}
with $\xi$ denoting the localization length, provided $\lambda$ is
taken large enough. Under these circumstances Eqs \eqref{bulk-Z-1}
- \eqref{bulk-Z-2} scale toward the critical ``edge" theory of the
quantum Hall state, Eq.\eqref{S-gap}, with $\sigma_H^\prime =
k(\nu) + \mathcal{O} (e^{-\lambda/\xi})$ now standing for the {\em
robustly} quantized Hall conductance, see Fig. \ref{FIG1}.
%
\begin{figure}
\includegraphics[width=100mm]{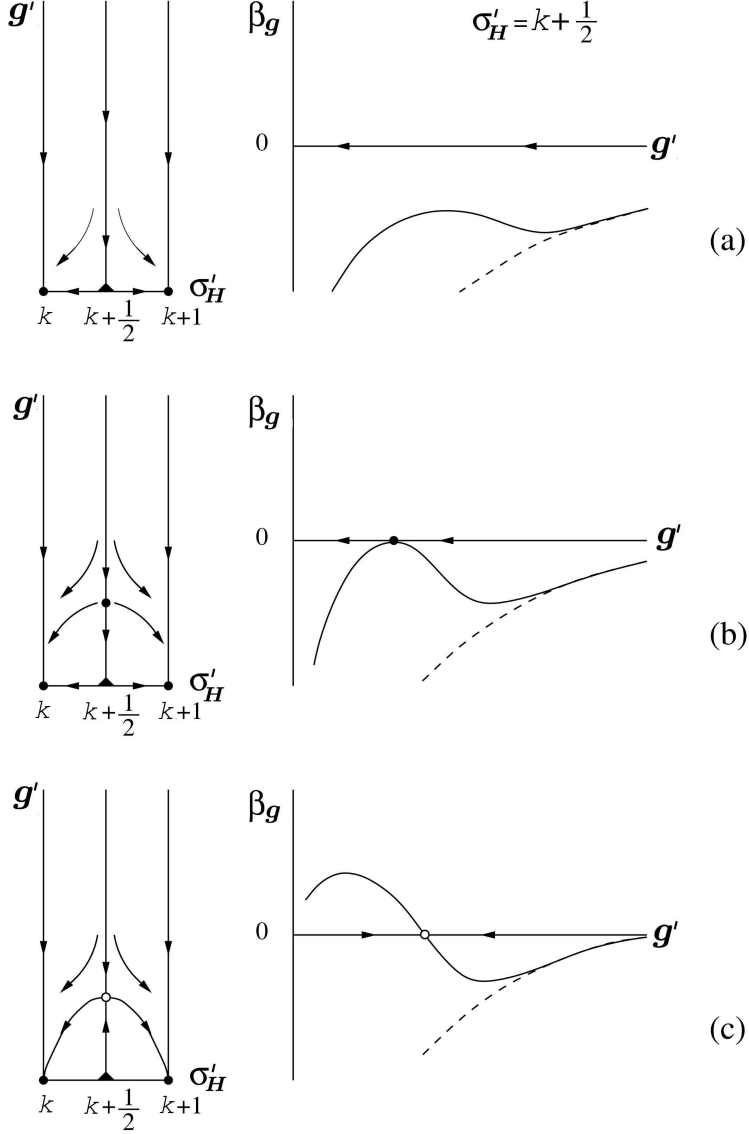}
\caption{ The renormalization group flow diagram in the $g^\prime$
- $\sigma^\prime_H$ ``conductance" plane, along with the
$\beta_{g} = d g^\prime / d \ln \lambda $ function along the line
$\sigma_{H}^\prime = k+\frac{1}{2}$, for different values of $M$
and $N$. The results show how instanton effects alter the
perturbative $\beta_g$ functions depicted by the dashed lines. (a)
Typical behavior for large values of $M$, $N$. (b) Intermediate
behavior that is likely displayed by the theory with $M=N=1$ or
the $O(3)$ non-linear $\sigma$ model. (c) Typical behavior for $0
\leqq M , N < 1$, see text.} \label{FIG3}
\end{figure}
%

Since nothing of the argument seems to crucially depend on the
number of field components $M$ and $N$, we expect that this
scaling picture has a quite general significance for the theory
for all values $M,N \ge 0$. Indeed, the explicit results for Eqs
\eqref{Kubo1} - \eqref{Hall-prime-1} obtained from the instanton
calculational technique \cite{PruiskenBurmistrov} are all in
accordance with the Thouless criterion and the super universality
concept discussed earlier, see Fig. \ref{FIG3}. On the other hand,
since the theory is generally inaccessible on the strong coupling
side, it is extremely important to have a simple example where the
different aspects of the super universality concept can be
investigated and explored exactly. For this purpose we specialize
from now onward to the large $N$ expansion of the $CP^{N-1}$
model.
\section{Weak coupling results at large $N$ \label{Observables}}
From Eqs \eqref{bulk-Z-1} - \eqref{Kubo2} we see that the physics
of the $\vartheta$ vacuum is in general defined by only three
physical quantities, namely the free energy $\mathcal{F}_b$ and
the response parameters $g^\prime$ and $\theta^\prime$. These
quantities are all defined by the underlying theory of bulk
excitations with a strictly quantized topological charge
$\mathcal{C} [ Q_0]$. This means that the $\theta (\nu)$
dependence can be expressed in terms of a series expansion in
``discrete topological sectors" $n$ according to
\be
 {(\lambda^\prime)^{-2}}\mathcal{F}_b =\sum_{n=0}^\infty \phi_n \cos n
 \theta(\nu)
 \label{FreeN-1}
\ee
and
\be
 g^\prime (\lambda^\prime) =g_0 + \sum_{n=0}^\infty \rho_n \cos n \theta(\nu)
 \label{g-prime-top}
\ee
\be
 \theta^\prime (\lambda^\prime) =\theta(\nu) + \sum_{n=1}^\infty \xi_n \sin n
 \theta(\nu).
 \label{theta-prime-top}
\ee
The $\phi_n$ , $\rho_n$ and $\xi_n$ generally stand for functions
of $g_0$ and the scale size $\lambda^\prime$ alone. As we shall
see in the remainder of this paper, it is precisely this feature
of the theory that eventually facilitates the contact between the
weak and strong coupling phases of the large $N$ expansion. We
first briefly recall, in Sections \ref{Instantons-N} and
\ref{Instantons-N-1} below, some of the results obtained in the
weak coupling instanton phase.\cite{PruiskenBurmistrov}
\subsection{Free energy \label{Instantons-N}}
The large $N$ expansion is usually expressed in terms of a
re-scaled parameter $g_0$
\be
 \sigma_0 = g_0 / N  \label{sigma1}
\ee
such that the perturbative quantum corrections to $\sigma_0$ are
independent of $N$
\be
 \sigma ( \lambda ) =  \sigma_0 - \frac{1}{2\pi} \ln
 \mu\lambda = - \frac{1}{2\pi} \ln \lambda M_0 . \label{sigma-corr-1}
\ee
Here, $M_0$ denotes the dynamically generated mass of the large
$N$ expansion
\be
 M_0 = \mu e^{-2\pi\sigma_0} \label{massM0}
\ee
Within the dilute instanton gas approach one usually deals with
only the $n=1$ term in Eq. \eqref{FreeN-1}. Using $\sigma$ as a
shorthand for $\sigma (\lambda)$ we cast the standard result for
large values of $N$ in the following general form
\be
 {(\lambda^\prime)^{-1}}\mathcal{F}_b =-\int \frac{d\lambda}{\lambda^3}
 w (\sigma) W^N (\sigma) \cos \theta (\nu) \label{FreeN}
\ee
where $(\lambda^\prime)^2$ denotes the area of the system and the
functions $W(\sigma)$ and $w(\sigma)$ are given by
\be \label{Instantonff0-0}
 W (\sigma) = \exp \left\{ - 2\pi \sigma + \ln 4\pi \sigma  -\gamma - \frac{1}{2}\right\}
 ~,~~ w (\sigma) = \left( {2N}/{\pi} \right)^{3/2} /e
\ee
with $\gamma\approx 0.577$ the Euler constant. The unspecified
integral over the scale size $\lambda$ in Eq. \eqref{FreeN}
diverges in the infrared. This notorious drawback dramatically
complicates the meaning of the semiclassical methodology.
\cite{Rajaraman,BergLuesher} Within the historical large $N$
analysis, for example, it was first assumed that instantons do not
contribute. \cite{Witten} Following the seminal critique by
Jevicki \cite{Jevicki} this conjecture was later abandoned.
\cite{Affleck80} The different approaches to the $\vartheta$
vacuum have nevertheless led to an ``arena of bloody controversies
" \cite{Coleman1} that has not been resolved even to date.

In what follows we shall argue that the results of Eq.
\eqref{Instantonff0-0} can only be trusted in the range $\lambda
M_0 \ll 1$ where one normally expects the perturbative quantum
theory to be valid. As one approaches the strong coupling phase
$\lambda M_0 \gtrsim 1$ the integrant of Eq. \eqref{FreeN}
presumably gets complicated by additional contributions from pairs
of instantons and anti-instanton that are difficult incorporate
semiclassically. In Section \ref{Pseudo-instanton} we show how the
large $N$ steepest descend methodology can be used to shed new
light on the matter. We find, in particular, that the general form
of Eq. \eqref{FreeN} is retained except that the functions
$W(\sigma)$, $w(\sigma)$ and $\sigma(\lambda)$ have a different
meaning when $\lambda M_0 \gtrsim 1$. These extended instanton
results are a special case of the more general statement which
says that the coefficients $\phi_n$ in Eq. \eqref{FreeN-1} are all
finite as $\lambda^\prime$ goes to infinity.

Even though the new insights into the traditional infrared
problems of instantons are important, free energy considerations
alone do not teach us much about the singularity structure of the
theory as $\theta(\nu)$ approaches $\pm\pi$. The lowest order
terms in the series of Eq. \eqref{FreeN-1} generally tell us
something about the regular part of the free energy which is of
secondary interest.
\subsection{Observable theory \label{Instantons-N-1}}
To study the low energy dynamics of the $\vartheta$ vacuum, in
particular the quantum Hall effect, one must develop a quantum
theory of the response parameters $g^\prime$ and $\theta^\prime$
in Eqs \eqref{g-prime-top} and \eqref{theta-prime-top} which we
term the {\em observable theory}.
The results can in general be expressed as an integral over
scale sizes $\lambda$
\begin{eqnarray}
 \sigma^\prime (\lambda^\prime) &=& \sigma^\prime (\lambda_0)
 + \int_{\lambda_0}^{\lambda^\prime} \frac{d\lambda}{\lambda}
 \beta_\sigma (\sigma^\prime , \theta^\prime)      \label{beta11}\\
 \theta^\prime (\lambda^\prime) &=& \theta^\prime (\lambda_0)
 + \int_{\lambda_0}^{\lambda^\prime} \frac{d\lambda}{\lambda}
 \beta_\theta (\sigma^\prime , \theta^\prime) \label{beta22}
\end{eqnarray}
where $\sigma^\prime = g^\prime/N$. The $\beta$ functions
extracted from the dilute instanton gas are universal and can be
written in the form
\begin{eqnarray}
 \beta_{\sigma} &=& f (\sigma^\prime) -
 f_1 (\sigma^\prime) ~W^N (\sigma^\prime) \cos \theta^\prime  \label{weakinstanton1} \\
 \beta_\theta &=&  ~~~~~~~-g_1 (\sigma^\prime)
 ~W^N (\sigma^\prime) \sin
 \theta^\prime   \label{weakinstanton2}
\end{eqnarray}
where the function $W$ is the same as in Eq.
\eqref{Instantonff0-0} and the $f$, $g$ functions are given by
\begin{eqnarray}
 f (\sigma^\prime) &=& -\frac{1}{2\pi\sigma^\prime} \nonumber \\
 f_1 (\sigma^\prime) &=& 8 e^{-1} N (2\pi N)^{1/2} ~\sigma^\prime \nonumber \\
 g_1 (\sigma^\prime) &=& 8 e^{-1} N (2\pi N)^{3/2} ~(\sigma^\prime)^2
 .\label{f-g-1}
\end{eqnarray}
In anticipation of our findings in Section \ref{Pseudo-instanton}
we can say that the general form of the $\beta$ functions in Eqs
\eqref{weakinstanton1} and \eqref{weakinstanton2}, like the free
energy of Eq. \eqref{FreeN}, is retained all the way down to the
strong coupling phase provided one gives a proper meaning to the
functions $W$, $f$ and $g$. These generalized instanton results
are the lowest order terms in an infinite series that reads
\begin{eqnarray}
 \beta_{\sigma} &=& \sum_{n=0}^\infty \beta_\sigma^{(n)} (\sigma^\prime)
 \cos n \theta^\prime  \label{all-instanton1} \\
 \beta_{\theta} &=& \sum_{n=1}^\infty \beta_\theta^{(n)} (\sigma^\prime)
 \sin n \theta^\prime  . \label{all-instanton2}
\end{eqnarray}
On the weak coupling side one expects that the convergence of the
series is controlled by instanton factors that are typically of
the form
\be
 \beta_\sigma^{(n)} (\sigma),~
 \beta_\theta^{(n)} (\sigma) \simeq W^{Nn} (\sigma^\prime)
 .\label{beta-n}
\ee
On the strong coupling side we find that the $\beta_\sigma$
function as a whole goes to zero and the quantities
$\beta_\theta^{(n)} (\sigma)$ in Eq. \eqref{all-instanton2} all
approach a finite constant in the limit $\sigma^\prime \rightarrow
0$. In different words, the large $N$ expansion is the much sought
after example where the concept of $\vartheta$ renormalization can
be explored and investigated in the entire range from weak to
strong coupling.
%
\begin{figure}[tbp]
\includegraphics[width=80mm]{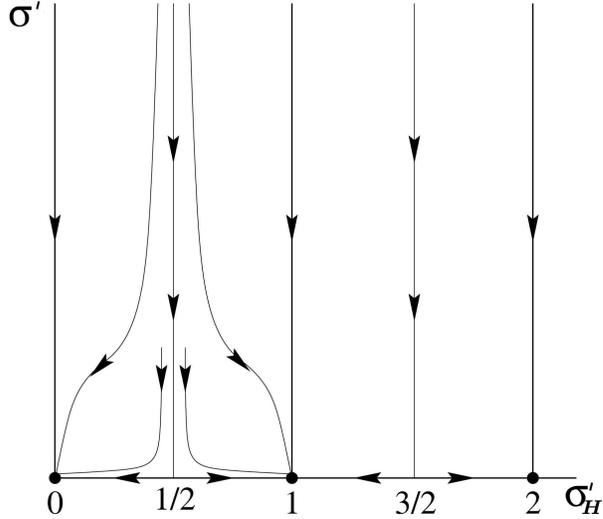}
\caption{ Renormalization group flows in the $\sigma^\prime$ -
$\sigma_H^\prime$ ``conductance" plane according to the $CP^{N-1}$
theory with large values of $N$, see text. } \label{FIG4}
\end{figure}
%
\section{$CP^{N-1}$ model\label{Sec4}}
\subsection{Introduction}
In this Section we review the various different steps of the large
$N$ steepest descend methodology of the $CP^{N-1}$
model.~\cite{Dadda,Witten,Affleck80} The matrix field variable $Q
\in {SU(N)}/{U(N-1)}$ can be expressed in terms of a complex
vector field $z_\alpha$
\be
 Q_{\alpha\beta} =
 2 z_\alpha^* z_\beta -\delta_{\alpha\beta}
\ee
with ${{z}^*_\alpha {z}_\alpha} = 1$. The action is usually taken
in $1+1$ space-time dimension and without mass terms
\be
 S [Q] = \int d^2 x\, \Bigl [ - g_0 \left( \partial_\mu
 z_\alpha^* \partial_\mu z_\alpha + z_\alpha^* \partial_\mu
 z_\alpha z_\beta^* \partial_\mu z_\beta \right)
 + \sigma_{H}^0 ~\varepsilon_{\mu\nu}
 \partial_\mu z_\alpha^* \partial_\nu z_\alpha \Bigr ].
 \label{SbCPN}
\ee
Introducing a vector field $A_\mu$ one can write
\be
 S [Q, A_\mu] = \int d^2 x\, \Bigl [ - g_0  | D_\mu z_\alpha |^2 + i
 \sigma_{H}^0  \varepsilon_{\mu\nu}
 \partial_\mu A_\nu \Bigr ] \label{largeN-gauge}
\ee
where $D_\mu = \partial_\mu +i A_\mu$ and the topological charge
becomes
\be
 \mathcal{C} [A_\mu]=\frac{1}{2\pi} \int d^2 x\,
 \varepsilon_{\mu\nu} \partial_\mu A_\nu  .\label{top-charge-Amu}
\ee
Notice that under the shift $A_\mu \rightarrow A_\mu + i
z_\alpha^* \partial_\nu z_\alpha$ the $A_\mu$ and $z_\alpha$
fields decouple and we obtain the original theory of Eq.
\eqref{SbCPN}. For the time being we shall ignore the problem of
the massless ``edge" excitations and assume, following the
historical papers, that $A_\mu$ is an unconstrained free field. We
lift the nonlinearity condition ${z^*_\alpha z_\alpha} =1$ as
usual introducing an auxiliary field $\phi (x)$
\be
 S [Q,A_\mu , \lambda] = S [Q,A_\mu] + i g_0 \int d^2 x\,
 \phi  \left( |z_\alpha |^2 -1 \right) .\label{lagrange}
\ee
The $z_\alpha$ vector fields are now free and can be eliminated.
This leads to an effective theory in terms of $\phi$ and $A_\mu$
fields alone
\be
 S[A_\mu, \phi] = - N \left[ \tr \ln(-D^2 +i\phi ) -i\sigma_0
 \int d^2 x ~ \phi \right] + i \sigma_{H}^0 \int d^2 x\, \varepsilon_{\mu\nu}
 \partial_\mu A_\nu .\label{SAL}
\ee
We have introduced the quantity $\sigma_0 = g_0 / N$ as before.
Eq. \eqref{SAL} has an $SU(N)$ invariant stationary point $i\phi
(x) = M_0^2$. Putting $A_\mu =0$ we obtain the following
expression for the mass $M_0$
\be
 \int \frac{d^2 k}{(2\pi)^2} \frac{1}{k^2 + M_0^2} = \sigma_0 .
\ee
Introducing an ultraviolet cutoff $\mu$
\be
 \sigma_0 = \frac{1}{2\pi} \ln \frac{\mu}{M_0}, \qquad M_0 =\xi^{-1} =
 \mu e^{-2\pi\sigma_0}
\ee
we obtain the same results as in Eqs \eqref{sigma1} and
\eqref{massM0} extracted from ordinary perturbation theory. Next,
by neglecting the fluctuations in the $\phi$ field which are of
order $N^{-1}$, we expand the theory as a power series in the
$A_\mu$ field. Assuming Lorentz invariance we obtain the standard
result
\be
 N \tr \ln (-D_\mu^2 +M_0^2 ) = N \tr \ln (-\partial_\mu^2 +M_0^2 )
 +\frac{N}{48\pi M_0^2} \int d^2 x\, F_{\mu\nu}^2
 \hspace{0.5cm}{}\label{F2munu}
\ee
and the effective action reads
\be
 S [A_\mu] = \int d^2 x\, \Bigl [ -\frac{N}{48\pi M_0^2}  F_{\mu\nu}^2 + i
 \sigma_{H}^0 ~ \varepsilon_{\mu\nu} \partial_\mu A_\nu \Bigr ] .\label{SAmu}
\ee
At this point the historical large $N$ methodology gets
complicated. The difficulties are immediately obvious when
considering the expression for the free energy
\be
 \mathcal{F} (\sigma_{H}^0) = - \ln \int D[A_\mu] e^{S[A_\mu]}
 = \frac{12 \pi \lambda^2 M_0^2}{N} (\sigma_{H}^0)^2 \label{vacF}
\ee
where
\be
 \lambda^2 = \beta L
\ee
denotes the area in space-time (from now onward we write $\lambda$
for $\lambda^\prime$ whenever convenient). Even though Eq.
\eqref{vacF} does not display any periodicity in $\sigma_H^0$, one
nevertheless argues on heuristic grounds that Eq. \eqref{vacF}
{\em should be} periodic. \cite{Witten,Affleck80} Based on the
electrodynamics ``picture" by Coleman, \cite{Coleman} for example,
one assumes that Eq. \eqref{vacF} is only correct in the interval
$-\frac{1}{2} < \sigma_H^0 \le \frac{1}{2}$. Outside this interval
it is energetically favorable for the system to materialize a pair
of ``charges" that move in opposite directions to the ``edges" of
the universe such as to maximally shield the background ``electric
field" $\sigma_{H}^0$. Eq. \eqref{vacF} should therefore be
expressed in terms of the {\em internally generated} ``electric
field" rather than the {\em bare} value $\sigma_H^0$. We will come
back to these ideas in Section \ref{Quantum-Hall}
\subsection{Bulk and edge excitations \label{quantumHalleffect}}
Although cleverly designed, Coleman's ad hoc arguments should not
be mistaken for an exact or complete theory of the $\vartheta$
vacuum. To demonstrate that something fundamental is missing we
employ the master formulae for the conductances of Section
\ref{Sec2}. Since the large $N$ methodology is manifestly $SU(N)$
invariant (there is no spontaneous symmetry breaking) the
insertion of a background matrix field $t(x)$ is immaterial and we
immediately obtain the trivial response
\be
 g^\prime (\omega) = \sigma^\prime_H (\omega)=0 \label{no-eHq}
\ee
in the limit $\omega=0$. This result conflicts with the quantum
Hall effect, in particular Eq. \eqref{qHe-edge-1}, indicating that
the massless chiral ``edge" excitations in the problem have been
overlooked. These edge excitations have disappeared in Eq.
\eqref{SAmu} the reason being that incorrect assumptions have been
made about the order in which the integrals over the $A_\mu$ and
$z_\alpha$ fields in Eq. \eqref{largeN-gauge} must be performed.

Guided by the analysis of Section \ref{Sec3} we next discuss the
subtle modifications in the large $N$ methodology that are
necessary in order to be able to extract the correct low energy
dynamics of the $\vartheta$ vacuum.

\begin{enumerate}
\item
First, in accordance with Eq. \ref{tQ0t} we split the vector field
$z_\alpha$ in ``edge" components $\varphi$ and ``bulk" modes
$\zeta$ according to
\begin{equation}
 z_\alpha = \sum_{\beta=1}^N  \varphi_{\alpha\beta}  \,
 \zeta_\beta
 \label{z-split} .
\end{equation}
Here, the vector field $\zeta_\alpha$ is constrained by the
boundary condition $\zeta_\alpha = e^{i\phi} \delta_{\alpha,1}$
with $\phi$ an arbitrary $U(1)$ gauge.

\item
Eq. \eqref{z-split} implies that the topological charge
$\mathcal{C} [A_\mu]$ in Eqs \eqref{largeN-gauge} and
\eqref{top-charge-Amu} must be split into integral and fractional
pieces. Specifically, we impose the constraint
\be
 \mathcal{C} [A_\mu] = n + \mathcal{C}[q] ~,~~
 \mathcal{C}[q] = \frac{1}{16\pi i} \oint d x_\mu \,
 \tr \, \epsilon_{\mu\nu} q \partial_\mu q \partial_\nu q   \label{constraint}
\ee
where $n$ is an arbitrary integer and $\mathcal{C}[q]$ the
fractional piece associated with the ``edge" component $\varphi$
or $q = \varphi^{-1} \Lambda \varphi$. Introducing an auxiliary
field $\eta$ we incorporate the constraint by substituting
\begin{eqnarray}
 i \sigma_H^0  \int d^2 x\,
 \varepsilon_{\mu\nu} \partial_\mu A_\nu
 &\rightarrow& i (\sigma_H^0  - \eta) \int d^2 x\,
 \varepsilon_{\mu\nu} \partial_\mu A_\nu  +
 2\pi i\eta (n+ \mathcal{C} [q]) \label{constraint-1}
\end{eqnarray}
in Eqs \eqref{largeN-gauge} and \eqref{top-charge-Amu}.

\item
The steps from Eq. \eqref{lagrange} to Eq. \eqref{F2munu} are
slightly modified since we only integrate over the ``bulk"
components $\zeta_\alpha$ while retaining the ``edge" matrix field
variable $\varphi$ or $q$. The results are summarized by making
the following substitution in Eqs \eqref{F2munu} and \eqref{SAmu}
\be
 \frac{N}{48\pi M_0^2} \int d^2 x\, F_{\mu\nu}^2
 \rightarrow \frac{N}{48\pi M_0^2} \int d^2 x\, F_{\mu\nu}^2
 + \delta S [q, A_\mu] .\label{Det+corr}
\ee
Here $\delta S$ stands for all the higher order terms that are
irrelevant. The only term of physical interest is the small
correction term
\be
 \delta S [q, A_\mu]
 = \frac{g^\prime (\lambda)}{8} \int d^2 x\,
 \tr \partial_\mu q \partial_\mu q ~,~~~~
 g^\prime (\lambda) \approx e^{- \lambda M_0} \label{delta-S}
\ee
that is permitted by the boundary condition imposed on the
$\zeta_\alpha$ vector field.
\end{enumerate}
By substituting Eqs \eqref{constraint-1} and \eqref{Det+corr} in
the historical result of Eq. \eqref{SAmu} we obtain a more complex
theory that besides the $A_\mu$ field also depends on the ``edge"
matrix field variable $q$ as well as the auxiliary field $\eta$
and $n$
\begin{eqnarray}
 S [A_\mu,q,\eta,n] &=& \int \Bigl [ -\frac{N}{48\pi M_0^2}  F_{\mu\nu}^2 + i
 (\sigma_H^0  - \eta) \varepsilon_{\mu\nu} \partial_\mu A_\nu\Bigr ]+
 \nonumber\\
 && ~~~~~~~~~~~~~~~~~~~~~~~~~~~~~~~
 + 2\pi i\eta (n+ \mathcal{C} [q]) +\delta S [q,A_\mu] .\label{SAmu-1}
\end{eqnarray}
At this stage of the analysis several remarks are in order. First
of all, it should be mentioned that the piece $\delta S$ in Eq.
\eqref{SAmu-1} really describes the correction terms in a
systematic expansion in large values of $N$. To see this we
re-scale
\be
 \lambda \rightarrow \lambda \sqrt{N} ~,~~
 A_\mu \rightarrow A_\mu / \sqrt{N}
\ee
while keeping the dimensionless quantities $q$, $\eta$ and $n$
unchanged. This removes the factor $N$ from the leading order
result in Eq. \eqref{SAmu-1} such that all the $N$ dependence now
appears in $\delta S$. For example, the quantity $g^\prime
(\lambda)$ gets replaced by
\be
 g^\prime (\lambda) \rightarrow
 g^\prime (\sqrt{N} \lambda) \approx e^{- \sqrt{N}
 \lambda M_0}
\ee
indicating that physically the large $N$ steepest descend
methodology describes a systematic expansion about the strong
coupling line $g^\prime =0$ in the $g^\prime$ - $\sigma_H^\prime$
conductance plane.

Secondly, from Eq. \eqref{delta-S} we infer that the natural
scaling parameter $\sigma^\prime (\lambda)$ of the large $N$
theory is given by
\be
 \sigma^\prime (\lambda) = g^\prime (\lambda) \approx e^{-
 \lambda M_0} \ll 1.\label{sigma-prime}
\ee
This definition is the strong coupling counter part of the weak
coupling statement of Eqs \eqref{sigma1} and \eqref{sigma-corr-1}.
Since Eq. \eqref{sigma-prime} does not depend on $\theta(\nu)$ we
will not distinguish, in what follows, between the observable
parameter $\sigma^\prime (\lambda)$ and the renormalized quantity
$\sigma (\lambda)$ in which the free energy is generally expressed
(see, however, Section \ref{relation-beta-sigma}).

Keeping these remarks in mind we discard, from now onward, the
piece $\delta S$ unless explicitly stated otherwise. We proceed by
eliminating the $A_\mu$ field in Eq. \eqref{SAmu-1} which is now a
free field. The effective action in terms of field variables
$\eta$, $q$ and $n$ reads
\begin{equation}
 S [\eta, q, n] = - \frac{12\pi \lambda^2 M_0^2}{N}
 (\sigma_H^0 - \eta)^2 + 2\pi i\eta (n+
 \mathcal{C} [q]) . \label{SbL1}
\end{equation}
The effective theory of edge excitations is now defined by
\be
 Z [q] = \sum_n \int_{-\infty}^\infty d\eta ~e^{S [\eta, q,
 n]}. \label{vacF-plus-q}
\ee
Notice that the only difference between Eqs \eqref{SbL1} and
\eqref{vacF-plus-q} and the original results of Eqs \eqref{SAmu}
and \eqref{vacF} is the constraint of Eq. \eqref{constraint} that
separates the integral topological sectors from the fractional
ones. In what follows we shall distinguish between two different
strong coupling phases of the theory, termed the {\em quantum Hall
phase} (Section \ref{Quantum-Hall}) and the {\em pseudo instanton
phase} (Section \ref{Pseudo-instanton}) respectively, depending on
the value of the dimensionless variable ${12\pi \lambda^2
M_0^2}/{N}$.
%
\begin{figure}
\includegraphics[width=80mm]{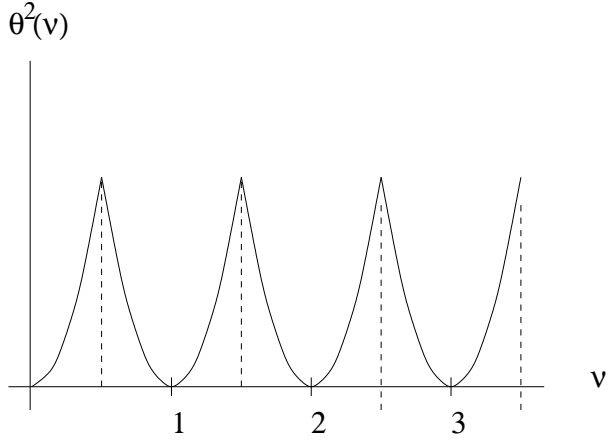}
\caption{ The bulk free energy $\mathcal{F}_b/\lambda^2 \propto
\theta^2 (\nu)$ with varying $\nu$, see text. } \label{FIG5}
\end{figure}
%
\section{Quantum Hall phase ${12\pi \lambda^2 M_0^2 }/{N} \gg 1$
\label{Quantum-Hall}}
By making use of the Poisson summation formula
\be
 \sum_{n} e^{2\pi i\eta n} = \frac{1}{2\pi} \sum_{m} \delta
 (\eta - m)
\ee
we can express Eqs \eqref{SbL1} and \eqref{vacF-plus-q} as sum
over integers $m$
\be \label{Zamu1}
 Z [q] = \sum_{m} \exp \left\{ -\frac{12 \pi \lambda^2  M_0^2}{N}
 (\sigma_H^0 - m)^2 +2\pi i m \mathcal{C} [q] \right\} .
\ee
This sum is rapidly converging in the limit where $\lambda
\rightarrow \infty$ while keeping all the other parameters in the
problem fixed. Let us next compare the completely different
results that are obtained depending on the interpretation of the
topological charge of the $A_\mu$ vector field.
\begin{enumerate}

\item
If one assumes, in accordance with the historical large $N$
analysis, that the original $A_\mu$ field is an unconstrained free
field then one must integrate the result of Eq. \eqref{Zamu1} over
the range $-\frac{1}{2} < \mathcal{C} [q] \leq \frac{1}{2}$. This
integral singles out the $m=0$ term in the series and the free
energy is exactly same as in Eq. \eqref{vacF}. We can write
\be
 \ln ~\langle Z [q] \rangle =
 \mathcal{F} (\sigma_H^0) =
 \frac{12 \pi \lambda^2 M_0^2}{N} (\sigma_{H}^0)^2 .\label{ZAmu-old}
\ee
where the brackets $\langle \dots \rangle \equiv
\int_{-\frac{1}{2}}^{\frac{1}{2}} d\mathcal{C} \dots $ denote the
average over the boundary conditions.

\item
If, on the other hand, we fix the boundary conditions on the
$A_\mu$ field or, as done in the present investigation, assign an
entirely different physical significance to the fractional piece
of the topological charge $\mathcal{C} [q]$  then the expression
of Eq. \eqref{Zamu1} is evaluated very differently. Employing the
split $\sigma_H^0 = \nu = \frac{\theta(\nu)}{2\pi} + k(\nu)$
introduced in Eq. \eqref{split3} and shifting the sum over $m$ we
obtain
\begin{eqnarray}
 Z [q ] &=& e^{2\pi i k(\nu) \mathcal{C} [q]} Z_b [q] \nonumber \\
 Z_b [q] &=& \sum_{m} \exp \left\{ -\frac{12 \pi \lambda^2  M_0^2}{N}
 (\frac{\theta(\nu)}{2\pi} - m)^2 +2\pi i m \mathcal{C} [q] \right\} .\label{Zamu2}
\end{eqnarray}
The sum in Eq. \eqref{Zamu2} is dominated by the $m=0$ term and
upon taking the thermodynamic limit $\lambda \rightarrow \infty$
we obtain
\be
 \ln Z [q] =  -\mathcal{F}_b (\theta(\nu)) + 2\pi i k(\nu) \mathcal{C}
 [q] \label{ZAmu3}
\ee
where $\mathcal{F}_b$ now denotes the free energy of the ``bulk"
which is given by
\be
 \mathcal{F}_b (\theta(\nu)) = \frac{3 \lambda^2 M_0^2}{\pi N} \theta^2
 (\nu)
 = \frac{12 \pi \lambda^2 M_0^2}{N} \left(\sigma_H^0 - k (\nu)\right)^2.
 \label{vacFb}
\ee
\end{enumerate}
Remarkably, Eqs \eqref{ZAmu3} and \eqref{vacFb} display all the
interesting physics of the $\vartheta$ vacuum that the historical
result of Eq. \eqref{ZAmu-old} did not give. Unlike Eq.
\eqref{ZAmu-old}, for example, the free energy of Eq.
\eqref{vacFb} is a periodic function of $\sigma_H^0 = \nu$ with a
sharp ``cusp" or first order phase transition at $\nu =
k+\frac{1}{2}$, see Fig. \ref{FIG5}. Eq. \eqref{vacFb} is in
accordance with Coleman's original ideas with $\theta(\nu) =
2\pi(\sigma_H^0 - k(\nu))$ standing for the {\em internally
generated} ``electric field" and $k(\nu)$ the part that originates
from the charges at ``edges" of the universe. In addition to this,
Eq. \eqref{ZAmu3} displays the quantum Hall effect. The piece
$2\pi i k(\nu) \mathcal{C} [q]$ is recognized as the action of
massless chiral edge excitations, see Eq.\eqref{S-gap}. The
integer $k(\nu)$ now stands for the {\em robustly} quantized Hall
conductance with sharp transitions occurring at half-integral
values of $\nu$, see Fig. \ref{FIG1}.

In conclusion, the correct physical interpretation of the large
$N$ theory crucially depends on a correct treatment of the
massless ``edge" excitations in the problem. By mishandling these
excitations like in Eq. \eqref{ZAmu-old} one actually looses all
the important ``bulk" phenomena and, consequently, one must work
very hard in order to retrieve at least some of the physics of the
$\vartheta$ angle. The historical papers on the subject did not
reveal the quantum Hall effect, however, nor did they provide a
correct physical understanding of issues like the quantization of
topological charge. \cite{Witten,Affleck80,Affleck83,Affleck91} We
will now proceed and investigate the consequences of our new
findings in more detail.
\subsection{Plateau transitions \label{PTs}}
First, to discuss the Thouless criterion introduced in Section
\ref{finite-size} we extend the result of Eq. \eqref{vacFb} to
include the effects of finite size scaling. By expanding the bulk
theory $Z_b$ to lowest order in $\mathcal{C} [q]$ and making use
of Eq. \eqref{delta-S} we obtain the more general expression
\be
 \ln Z [q] \rightarrow  -\mathcal{F}_b
 + 2\pi i \left( k(\nu) + \frac{\theta^\prime (\lambda)}{2\pi} \right) \mathcal{C}
 [q] - \frac{g^\prime (\lambda)}{8} \int d^2 x\, \partial_\mu q \partial_\mu q
 \label{vacFb-1}
\ee
where for $\theta(\nu) \approx 0$ the response parameters are
given by
\be
 g^\prime (\lambda) = \sigma^\prime (\lambda)\backsimeq e^{-\lambda M_0} ~~,~~~
 \theta^\prime (\lambda) \backsimeq e^{-{12\pi
 \lambda^2 M_0^2}/{N}} .\label{qHe-corrections-1}
\ee
These results are precisely the same as $\tilde{S}^\prime_\sigma
[q]$ defined in Eq. \eqref{Bulk-prime-2}. The corrections in Eq.
\eqref{qHe-corrections-1} are slightly different from the naive
expectations of Eq. \eqref{qHe-corrections} based on exponential
localization. This difference is due to that fact that all the
interesting physics of the large $N$ theory, unlike the electron
gas, occurs along the strong coupling line $g^\prime (\lambda)
=0$.

\begin{figure}
\includegraphics[width=80mm]{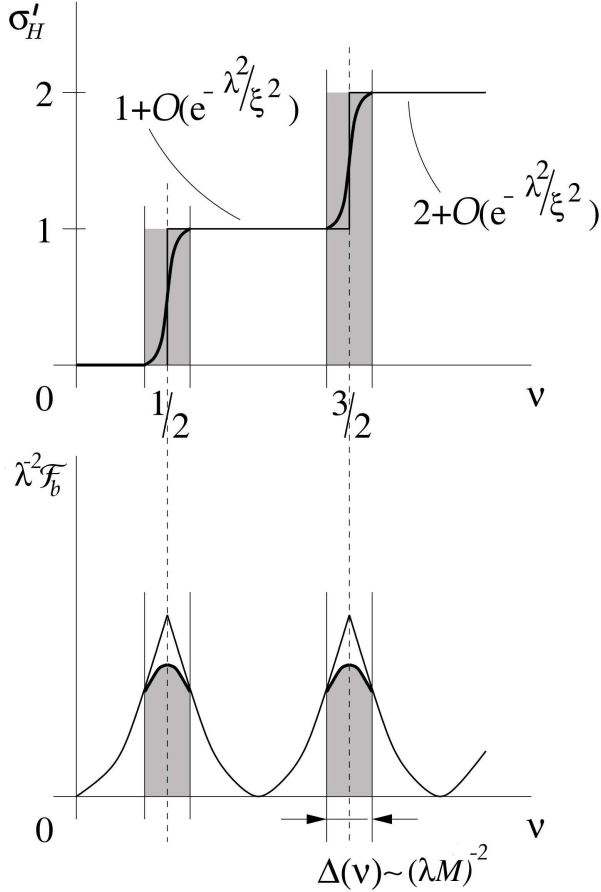}
\caption{Finite size scaling of the Hall conductance $\sigma'_{H}$
and the bulk free energy $\mathcal{F}_b /\lambda^2$ with varying
$\nu$. The dark areas at $\nu = 1/2, ~ 3/2$ ~etc. indicate the
critical regimes $\Delta (\nu) \propto \lambda^{-2}$ where
$\sigma^\prime_H$ varies smoothly from one plateau value to the
next. Similarly, the ``cusp" in $\mathcal{F}_b /\lambda^2$ is
smoothed out due to the finite scale size $\lambda$, see text. }
\label{FIG6}
\end{figure}

Next, to address the quantum Hall plateau {\em transitions} we
notice that when $\theta(\nu)$ approaches $\pm\pi$ the series of
Eq. \eqref{Zamu2} is dominated by the terms with $m=0,1$ and
$m=0,-1$ respectively. By expanding the bulk theory $Z_b$ to
lowest order in $\mathcal{C} [q]$ we obtain the same general
expression as in Eq. \eqref{vacFb-1} but with the following
scaling result for $\theta^\prime (\lambda)$
\be \label{theta-0}
 {\theta^\prime} (\lambda) = \pm {2\pi} \frac{e^{-2X}}{1+e^{-2X}}
\ee
where $\pm$ denotes the sign of $\theta(\nu)$. The scaling
variable $X$ is given by
\be
 X =  \frac{12\pi \lambda^2 M_0^2}{N}  \left ( \frac{1}{2} - \left| \frac{\theta (\nu)
 }{2\pi } \right| \right ) .
 \label{Xold}
\ee
The quantum Hall plateau transitions therefore display all the
characteristics of a continuous phase transition with a diverging
correlation length $\xi$
\be
 X = \frac{\lambda^2}{2 \xi^2} , ~~\xi = \frac{\sqrt{N}}{4 \sqrt{3\pi} M_0} \left( 1 - \left|
 \frac{\theta(\nu)}{\pi} \right| \right)^{- \frac{1}{2}} .
 \label{div-xi}
\ee
The finite size scaling behavior of the Hall conductance
$\sigma_{H}^\prime (\lambda) = k(\nu) + \frac{\theta^\prime
(\lambda) }{2\pi}$ and the free energy $\mathcal{F}_b$ with
varying values of the filling fraction $\nu$ is illustrated in
Fig. \ref{FIG6} and will be discussed further in Section
\ref{cond-distr}. It is interesting to notice that the scaling
results for the quantum Hall plateau transitions are essentially
the same as those taken from the free electron gas
\cite{Pruisken88} and observed in the experiment.\cite{Wei}
%
%
\subsection{$\vartheta$ renormalization \label{theta-ren}}
\begin{figure}
\includegraphics[width=80mm]{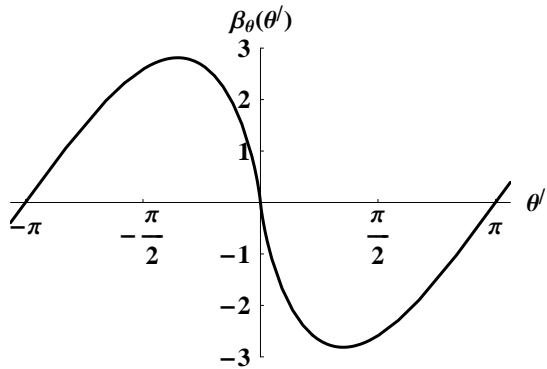}
\caption{The function $\beta_\theta$ with varying $\theta^\prime$,
see text. } \label{FIG7}
\end{figure}

From Eqs \eqref{sigma-prime} and \eqref{theta-0} we obtain the
$\beta$ functions (see also Fig. \ref{FIG7})
\begin{eqnarray}
 \beta_\sigma (\sigma^\prime) &=& \frac{d \sigma^\prime}{d\ln \lambda} =
 \sigma^\prime \ln \sigma^\prime \label{beta-sigma}\\
 \beta_\theta (\theta^\prime) &=& \frac{d \theta^\prime}{d\ln \lambda}
 = \frac{\theta^\prime}{\pi} \Bigl [ 2\pi -
 |\theta^\prime| \Bigr ]
 \ln \Bigl [ \frac{|\theta^\prime|}{2\pi - |\theta^\prime|} \Bigr ]
 \label{beta-theta}
\end{eqnarray}
which are amongst the most important results of this paper. Eqs
\eqref{beta-sigma} and \eqref{beta-theta} together with the weak
coupling instanton results of Section \ref{Instantons-N-1} give
rise to the renormalization group flow lines of Fig. \ref{FIG4}.
We identify two different kinds of strong coupling fixed points, a
massive one at $\sigma^\prime = \theta^\prime=0$ and a critical
one at $\sigma^\prime= 0$ and $\theta^\prime = \pm\pi$. Near
$\theta^\prime=0$ we have
\be
 \beta_\sigma = \beta_\sigma (\sigma^\prime) =
 \sigma^\prime \ln \sigma^\prime ~,~~
 \beta_\theta = \beta_\theta (\theta^\prime) = 2 {\theta^\prime}
 \ln |\theta^\prime| \label{beta-theta-1}
\ee
indicating that the Hall conductance is robustly quantized. Near
$\theta^\prime =\pi$ for $\sigma^\prime = 0$ we find
\be
 \frac{d (\pi - \theta^\prime)}{d\ln \lambda} = 2 (\pi -
 \theta^\prime) .\label{exponent-2}
\ee
The exponent value $2$ is the inverse of the correlation length
exponent given by Eq. \eqref{div-xi} and, at the same time, a
standard result for a first order phase transition in two
dimensions.
%
\subsection{Conductance distributions \label{cond-distr}}
A very interesting feature of the large $N$ expansion is that
$Z[q]$ near the plateau transition can be evaluated exactly and
not just to lowest orders in a series expansion in $\mathcal{C}
[q]$. The exact result is most conveniently expressed in terms of
the Hall conductance $\sigma_{H}^\prime (\lambda) = k(\nu) +
\frac{\theta^\prime (\lambda)}{2\pi}$. Let $\nu \approx k_0 +
\frac{1}{2}$ denote the transition regime with $k_0$ an arbitrary
integer. We then obtain
\begin{eqnarray}
 && Z [q] = e^{-\mathcal{F}_b } \times
 \left\{ (1+k_0 - \sigma_{H}^\prime ) e^{2\pi i k_0 \mathcal{C} [q]} +
 (\sigma_{H}^\prime -k_0) e^{2\pi i (k_0 +1)\mathcal{C} [q]}
 \right\}~ \label{k0-k0plus1}
\end{eqnarray}
where
\be
 \sigma_{H}^\prime =\sigma_{H}^\prime (\lambda) = k_0 + \frac{1}{2} e^{\lambda^2 M^2
 \Delta \nu} \cosh^{-1} (\lambda^2 M^2 \Delta \nu) ~~,~~~\Delta \nu = \nu-k_0 -
 \frac{1}{2} .
 \label{k0-k0plus1xy}
\ee
Notice that Eq. \eqref{k0-k0plus1xy} varies continuously from one
plateau value ($k_0$) to the next ($k_0 +1$) as the relative
filling fraction $\Delta \nu$ varies from negative to positive
values, see Fig. \ref{FIG6}. On the other hand, Eq.
\eqref{k0-k0plus1} describes this transition in terms of a
probability distribution of the quantum Hall states $k_0$ and $k_0
+1$ that are represented by the phase factors $e^{2\pi k_0 i
\mathcal{C} [q]}$ and $e^{2\pi (k_0 +1) i \mathcal{C} [q]}$
respectively.

In order to see that Eq. \eqref{k0-k0plus1} actually defines a
{\em distribution} of the fractional part $\theta^\prime
(\lambda)$ of the Hall conductance we introduce the set of
variables
\begin{equation}
 \theta_n = 2\pi(k_0 -k(\nu) + n)
\end{equation}
with $n=0,1$. The bulk part $Z_b [q]$ in Eq. \eqref{k0-k0plus1}
can now be written as a sum
\be
 Z_b [q] = e^{-\mathcal{F}_b } \times \sum_{n=0}^{1}
 P_n  ~ e^{i\theta_n \mathcal{C} [q]} .
 \label{k0-k0plus1a}
\ee
Here, $P_n = P_n (\theta^\prime (\lambda))$ is a normalized weight
\be
 \sum_{n=0}^1 P_n =1~,~~~
 \left( P_{0} , ~P_{1} \right) =
 \left( 1 + \frac{\theta_0 - \theta^\prime (\lambda)}{2\pi} , 1 -
 \frac{\theta_1 - \theta^\prime (\lambda)}{2\pi} \right)
 \label{P-k0-k0plus1}
\ee
and the expectations are denoted by
\be
 \langle \theta^k \rangle = \sum_{n=0}^1 P_n \theta^k_n ~,~~ \langle \theta \rangle
 = \theta^\prime (\lambda).
\ee
These results indicate that the bulk quantity $\theta^\prime
(\lambda)$ is actually broadly distributed in a highly
non-gaussian manner. For example, we can express Eq.
\eqref{k0-k0plus1a} in terms of a cumulant expansion
\be
 \ln Z_b [q] = -\mathcal{F}_b +  \tilde{S}_\sigma^\prime [q] \label{Fb-Seff}
\ee
where
\be
 \tilde{S}_\sigma^\prime [q] =  i \theta^\prime (\lambda) \mathcal{C} [q] -
 \frac{1}{2\mathcal{!}} \zeta_2^\prime (\lambda) \mathcal{C}^2 [q] -
 \frac{i}{3\mathcal{!}} \zeta_3^\prime (\lambda) \mathcal{C}^3 [q] + \mathcal{O}
 (\mathcal{C}^4) .
 \label{Seff-k0-k0plus1}
\ee
The higher order cumulants multiplying $\mathcal{C}^2 [q]$,
$\mathcal{C}^3 [q]$ etc. can all be expressed in terms of the
``averaged" quantity $\langle \theta \rangle
 = \theta^\prime (\lambda)$. For example
\be
 \zeta_2^\prime (\lambda) = \langle \theta^2 \rangle - \langle \theta \rangle^2 =
 |\theta^\prime (\lambda)|(2 \pi - |\theta^\prime (\lambda) |).
 \label{Seff-k0-k0plus1a}
\ee
Notice that as one approaches the transition $\theta^\prime
(\lambda) \rightarrow \pi$ the ``root mean square fluctuation"
$\sqrt{\zeta_2^\prime (\lambda)}$ becomes equal to the averaged
value $\pi$. Away from the transition both the averaged value
$\theta^\prime (\lambda) $ and the higher order cumulants
$\zeta^\prime (\lambda) $ become exponentially small in $\lambda$.
The large $N$ expansion is therefore a prototypical example of the
broad ``mesoscopic conductance distributions" that are of interest
in the quantum theory of metals. \cite{CohenPruisken}
Distributions like Eq. \eqref{P-k0-k0plus1} do not affect the
scaling behavior of the system, however, since this behavior
depends on the ``ensemble averaged" quantity $\theta^\prime
(\lambda)$ alone.

\subsection{Twisted boundary conditions \label{twist}}
In the original papers in the field~\cite{Pruisken84} it was
already argued on general grounds that {\em de-}localized or
extended ``bulk" excitations at $\theta (\nu) = \pm \pi$ must
generally exist for all values $M,N \geq 0$. The idea naturally
emerges from 't Hooft's duality argument that is based on the
response of the system to imposing twisted boundary conditions.
The effect of these boundary conditions is obtained by inserting
$\mathcal{C} [q] = \frac{1}{2}$ in Eq. \eqref{k0-k0plus1a}. We can
write
\be
 -\ln Z_b [q] = \mathcal{F}_b -\ln \sum_{n=0}^{1}
 P_n (\theta^\prime) ~ e^{i\theta_n /2} = \mathcal{F}_b + \Delta \mathcal{F}.
\ee
The shift $\Delta \mathcal{F}$ in the free energy due to twisted
boundary conditions can be expressed in terms of $\theta^\prime$
defined in Eq. \eqref{theta-0} and the result is
\be
 \Delta \mathcal{F} (\theta^\prime ) =
 -\ln\left( 1- \frac{|\theta^\prime|}{\pi} \right) .\label{tHooft}
\ee
As long as $\theta(\nu)$ is different from $\pm \pi$ the shift
$\Delta \mathcal{F}$ is exponentially small in the scale size
$\lambda$ indicating that the system has a mass gap. However, when
$\theta(\nu)$ approaches $\pm \pi$ the response $\Delta
\mathcal{F} \propto \ln \xi$ diverges which means that the system
now has gapless ``bulk" excitations. Notice that these findings
are entirely consistent with all the other ideas and results that
have been discussed sofar, in particular Coleman's picture of
dissociating charges at $\theta(\nu) =\pm \pi$, the scaling
functions describing the quantum Hall plateau transitions as well
as the statistics of conductance distributions addressed in the
previous section.

In a subsequent paper we will embark on the critical correlations
of the large $N$ theory and show that they map onto the one
dimensional Ising model at zero temperature.
\cite{PruiskenBurmistrovShankar} This is unlike the grassmannian
theory with $0 \le M,N \lesssim 1$ where, as well is known, the
transition is a second order one with exponents that vary
continuously with varying $M$ and $N$. \cite{PruiskenBurmistrov}
Despite these and many other differences, the basic phenomena of
{\em de}-localization and scaling, including the robust
quantization of the Hall conductance, are nevertheless the same.
\section{Pseudo instanton phase ${12\pi \lambda^2 M_0^2}/{N} \ll 1$
and cross-over\label{Pseudo-instanton}}
We have now completed the strong coupling quantum Hall side of the
large $N$ expansion. In this Section we embark on the problem of
cross-over between the weak coupling instanton phase discussed in
Section \ref{Observables} and the strong coupling results of the
previous Section. For this purpose we consider Eqs \eqref{SbL1}
and \eqref{vacF-plus-q} in the regime where the dimensionless
quantity ${12\pi \lambda^2 M_0^2}/{N}$ is small. To obtain a
rapidly converging series we simply perform the integral over the
auxiliary field $\eta$ such that the theory be written as sum over
integral topological sectors $n$
\begin{eqnarray}
 Z [q ] &=& e^{2\pi i k(\nu) \mathcal{C} [q]} Z_b [q] \nonumber \\
 Z_b [q] &=& \sum_{n} \exp \left\{ -\kappa (\lambda)
 (n+\mathcal{C} [q])^2 + i \theta (\nu ) (n+ \mathcal{C} [q]) \right\}
 .\label{Zb}
\end{eqnarray}
where
\be
 \kappa (\lambda) = \frac{\pi N}{12 \lambda^2 M_0^2} \gg 1 .
 \label{kappa-sigma}
\ee
Notice that there exists a large regime in $\lambda$ where the sum
over $n$ is rapidly converging and, at the same time, the
condition for strong coupling $\lambda M_0 \gg 1$ is satisfied. In
terms of the scaling parameter $\sigma (\lambda)$ or
$\sigma^\prime (\lambda)$ we can write
\be
 e^{-\sqrt{12N/\pi}} \ll \sigma (\lambda) = \sigma^\prime (\lambda) =
 e^{-\lambda M_0} \ll 1 .
 \label{sigma-sigma}
\ee
We term this regime the {\em pseudo instanton phase} because Eq.
\eqref{Zb} defines a trigonometric series in $\theta(\nu)$ which
is in many ways similar to the one obtained from the semiclassical
theory based on instantons (Section \ref{Observables}). By
expanding Eq. \eqref{Zb} in powers of $\mathcal{C}[q]$ we obtain
the same general form as before
\be
 \ln Z_b [q] = -\mathcal{F}_b + i \theta^\prime ~ \mathcal{C} [q] + \dots
 \label{Fb-Seff-dots}
\ee
In what follows we separately consider the free energy
$\mathcal{F}_b$ (Section \ref{Free-final}) and the observable
theory $\theta^\prime$ (Section \ref{Pseudo-observables})
respectively.
\subsection{Free energy \label{Free-final}}
The free energy can be expressed in the general form of Eq.
\eqref{FreeN-1}
\be
 \lambda^2 {\mathcal{F}_b } =
 - \frac{24 M_0^2}{\pi N} \sum_{n=0}^{\infty} \phi_n (\kappa) \cos
 n\theta(\nu) . \label{F-strong1-a}
\ee
Here, the functions $\phi_n (\kappa)$ to lowest orders in $n$ are
computed to be
\begin{eqnarray}
 ~~~~~~~~~~~~~~~~~~~~~~~
 \phi_0 (\kappa) &=& -\frac{1}{2} \kappa e^{-2\kappa} + \mathcal{O}(e^{-4\kappa})
 \nonumber\\
 \phi_1 (\kappa) &=& ~~~~ \kappa e^{-\kappa} ~~+ \mathcal{O}(e^{-3\kappa})
 \nonumber\\
 \phi_2 (\kappa) &=& -\frac{1}{4} \kappa e^{-2\kappa} +
 \mathcal{O}(e^{-4\kappa}).\label{phi-012}
\end{eqnarray}
More generally one finds the dominating behavior $\phi_n (\kappa)
\simeq e^{-n\kappa}$ for $n> 0$. By expressing $\kappa$ in terms
of $\sigma$ given in Eq. \eqref{sigma-sigma} one can write this
result as $\phi_n (\kappa) \simeq \left[ \exp\{-\frac{\pi}{12
\ln^2 \sigma}\} \right]^{nN}$. This behavior is reminiscent of the
instanton factors $W^{nN} (\sigma)$ found in the weak coupling
regime (Section \ref{Observables}).
\subsubsection{Relation to quantum Hall phase}
Let us first establish the contact with the quantum Hall phase.
For this purpose we go back to the result of Eq.~\eqref{vacFb}
which is the $\lambda = \infty$ limit of the free energy. This
result can be expanded in terms of a trigonometric series
according to
\begin{eqnarray}
 ~~~~~~~~~~~~~~~~~~~
 \lambda^{-2} {\mathcal{F}_b } &=& ~~~\frac{3 M_0^2}{\pi N} ~\theta^2
 (\nu) \nonumber\\
 &=&  - \frac{24 M_0^2}{\pi N} \sum_{n=0}^{\infty} \phi_n (0) \cos
 n\theta(\nu)  . \label{F-strong1}
\end{eqnarray}
The coefficients $\phi_n (0)$ are defined by
\begin{eqnarray}
 \phi_n (0) &=& - \frac{1}{8\pi} \int_{-\pi}^{\pi} \vartheta^2 \cos(n\vartheta)
 {d\vartheta} = (-1)^{n+1} \frac{1}{2 n^2}
 ,\qquad n>0  \label{coeff.} \\
 \phi_0 (0) &=& -\frac{1}{16\pi} \int_{-\pi}^{\pi} \vartheta^2 {d\vartheta} =
 - \frac{\pi^2}{24}.
\end{eqnarray}
It is clear that the coefficients $\phi_n (0)$ in Eq.
\eqref{F-strong1} are the $\kappa = 0$ limit of the functions
$\phi_n (\kappa)$ in Eq. \eqref{F-strong1-a}. Eq.
\eqref{F-strong1} is retained also if one takes the finite size
scaling corrections into account, see Fig.\ref{FIG6}. The only
difference is that the coefficients $\phi_n (0)$ in Eq.
\eqref{F-strong1} are replaced by a series in powers of $\kappa$.
%
\begin{figure}
\includegraphics[width=80mm]{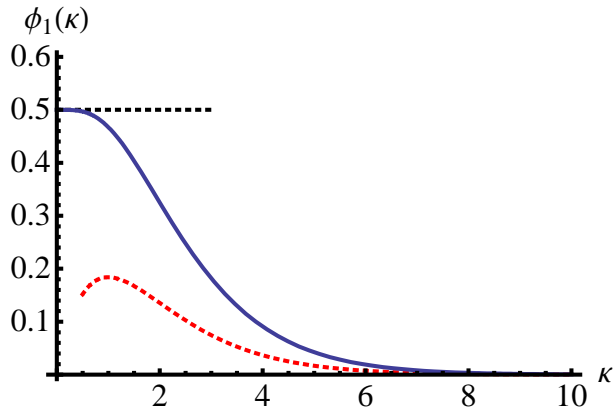}
\caption{The function $\phi_1 (\kappa)$ with varying $\kappa$, see
text.} \label{FIG8}
\end{figure}
%
As a typical example of the cross-over between the quantum Hall
and pseudo instanton phases we consider the function $\phi_1
(\kappa)$. A detailed investigation of this function in the
quantum Hall phase shows that the series expansion in $\kappa$ is
of the form
\begin{eqnarray}
 \phi_1 (\kappa) &=& \frac{1}{2} + \sum_{s=1}^\infty a_s \kappa^{2s}
 ~;~~~~~~~~~ \kappa \ll 1 \label{varphi-0}
\end{eqnarray}
where the lowest order coefficient $a_1$ in the series is
negative. On the other hand, from Eq. \eqref{phi-012} we obtain
the result for the pseudo instanton phase
\begin{eqnarray}
 \phi_1 (\kappa) &=& \kappa e^{-\kappa} ~;~~~~~~~~~~~~~~~~~~~~\kappa \gg 1
 \label{varphi-1}
\end{eqnarray}
From Eqs \eqref{varphi-0} and \eqref{varphi-1} we infer that
$\phi_1 (\kappa)$ can in general be written as the product of an
algebraic part $v_1 (\kappa)$ and an exponential part
$e^{-\kappa}$
\be
 \phi_1 (\kappa) = v_1 (\kappa) e^{-\kappa} .\label{v-1}
\ee
For the purpose of illustration we employ the following trial
function that satisfies the asymptotic constraints of Eqs
\eqref{varphi-0} and \eqref{varphi-1}
\be
 \phi_1 (\kappa) = \frac{1}{2} \sqrt{1+\kappa^2} \exp
 \left\{-\sqrt{1+\kappa^2} + 1 \right\} .\label{trial-f1}
\ee
This function along with the asymptotic behavior of Eqs
\eqref{varphi-0} and \eqref{varphi-1} is plotted in Fig.
\ref{FIG8}. Eq. \eqref{trial-f1} determines the function $v_1
(\kappa)$ in Eq. \eqref{v-1} to be
\be
 v_1 (\kappa) = \frac{1}{2} \sqrt{1+\kappa^2} \exp
 \left\{-\sqrt{1+\kappa^2} + 1 +\kappa \right\} . \label{trial-f1-0}
\ee
This function is algebraic in the sense that it smoothly
interpolates between $v_1 = \frac{1}{2}$ for small $\kappa$ and
$v_1 \propto \kappa$ for large values of $\kappa$.
\subsubsection{Relation to instanton phase}
Like the dilute instanton gas result discussed in Section
\ref{Instantons-N}, the free energy of the pseudo instanton phase
is dominated by the $n=1$ term in Eqs \eqref{F-strong1-a} and
\eqref{phi-012}. Given the general form of $\phi_1 (\kappa)$ in
the strong coupling phase, Eq. \eqref{v-1}, it is not difficult to
cast this term in the typical form of the dilute instanton gas.
More specifically, let $\lambda^\prime$ denote the system size
then we can write the result as an integral over scale $\lambda$
as follows
\begin{eqnarray}
 ~~~~~~~~~~~~~~~~~
 {(\lambda^\prime)^{-2}}{\mathcal{F}_b } &\approx&
 - \frac{24 M_0^2}{\pi N} \phi_1 (\kappa (\lambda^\prime)) \cos \theta(\nu) \nonumber \\
 &=& ~-\int_0^{\lambda^\prime} \frac{d\lambda}{\lambda^3}
 ~~w (\lambda) W^N (\lambda) \cos\theta (\nu) ~~~~\label{intF0}
\end{eqnarray}
where the functions $w$ and $W$ are given by
\be \label{W-strong-1}
 w (\lambda) = 4\left( \frac{d v_1 (\kappa)}{d\kappa}  - v_1 (\kappa) \right)~,~~
 W (\lambda) = \exp(-\frac{\pi}{12 \lambda^2 M_0^2})
\ee
with $v_1 (\kappa)$ an algebraic function of $\kappa = \kappa
(\lambda)$.
\begin{figure}
\includegraphics[width=80mm]{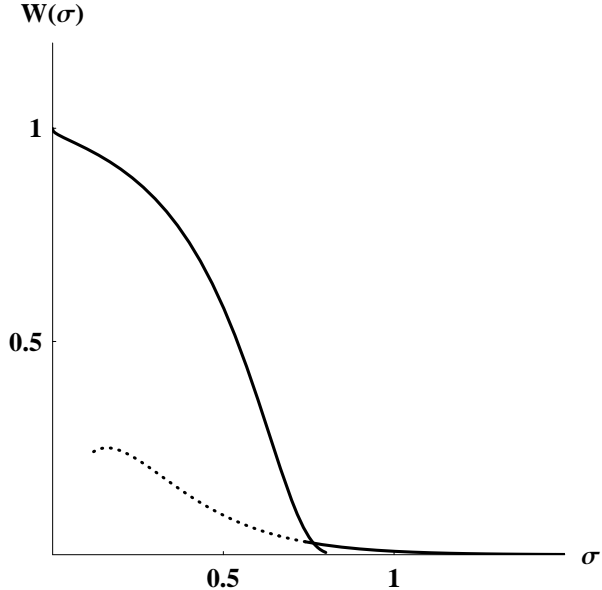}
\caption{ The function $W (\sigma)$ with varying $\sigma$
describing the cross-over between and the dilute instanton gas at
$\sigma \rightarrow \infty$ and the quantum Hall phase at $\sigma
\rightarrow 0$. The two different branches are the weak coupling
instanton result of Eq. \eqref{Instantonff0-0} and the strong
coupling result of Eq. \eqref{W-w-strong} respectively, see text.}
\label{FIG9}
\end{figure}
%
The contact with the instanton result of Eq.
\eqref{Instantonff0-0} is complete if we express the functions
$W(\lambda)$ and $w(\lambda)$ in terms of the scaling parameter
$\sigma(\lambda) = e^{-\lambda M_0} \ll 1$ rather than the scale
size $\lambda$. The strong coupling expressions for the
$w(\sigma)$ and $W(\sigma)$ functions read
\begin{equation}
 w (\sigma) = 4\left( \frac{d v_1 (\kappa (\sigma))}{d\kappa(\sigma)}
 - v_1 (\kappa(\sigma)) \right)~,~~
 W(\sigma) = \exp(-\frac{\pi}{12 \ln^2 \sigma})
 ~,~~\sigma \ll 1 \label{W-w-strong}
\end{equation}
where $\kappa(\sigma) = {\pi N}/{12 \ln^2 \sigma}$.
\subsubsection{Instantons regained}
In conclusion, the large $N$ steepest descend methodology and the
semiclassical instanton methodology are complementary descriptions
that together elucidate the complete structure of the $\vartheta$
vacuum. We find, in particular, that semiclassical concepts like
the ``dilute instanton gas" and ``discrete topological sectors"
have a universal significance that extends all the way down to the
strong coupling phase of the large $N$ expansion. By comparing Eqs
\eqref{W-w-strong} and \eqref{Instantonff0-0}, for example, one
sees that the cross-over between the weak and strong coupling
phases is primarily controlled by a single, $N$-independent
function $W(\sigma)$. This function is monotonically decreasing
from unity to zero as $\sigma$ increases from zero to infinity,
see Fig. \ref{FIG9}. On the other hand, the results of Eqs
\eqref{trial-f1-0} and \eqref{W-strong-1} clearly show that main
task of the $w(\sigma)$ function is to correctly describe the
cross-over from the pseudo instanton phase to the strong coupling
quantum Hall phase where the factor $W^N$ in Eq. \eqref{intF0} is
close to unity. Generally speaking, however, the $w(\sigma)$
function is a $1/N$ correction relative to $W(\sigma)$ and
therefore of secondary importance.

The results for the $W(\sigma)$ function plotted in Fig.
\ref{FIG9} furthermore indicate that the semiclassical theory of
instantons described by Eq. \eqref{Instantonff0-0} is only valid
in the range $\sigma \gtrsim 1$ or $\lambda M_0 \lesssim 1$ as
expected. Upon entering the strong coupling phase the functions
$W(\sigma)$ and $w(\sigma)$ generally have a different meaning
which cannot be obtained from semiclassical arguments alone. The
important conclusion to be drawn from the present analysis is that
the thermodynamic limit of the dilute instanton gas truly exists
and is finite.
\subsection{Physical observables \label{Pseudo-observables}}
Next, we embark on the observable quantity $\theta^\prime$ defined
in Eq. \eqref{Fb-Seff-dots}. By taking only the lowest order terms
$n=0, \pm 1$ in Eq. \eqref{Zb} into account we obtain
\begin{eqnarray}
 \theta^\prime (\lambda) &=& \theta(\nu) ~~ -  4\kappa(\lambda)
 e^{-\kappa(\lambda)} \sin \theta (\nu) .
 \label{S-edge-theta}
\end{eqnarray}
By substituting $\sigma^\prime = \sigma^\prime (\lambda)
=e^{-\lambda M_0}$ for $\kappa(\lambda)$ we obtain the following
renormalization group equations for the pseudo instanton phase
\begin{eqnarray}\label{pseudo-beta-sigma}
 \beta_\sigma &=& \frac{d\sigma^\prime}{d\ln\lambda} = \sigma^\prime \ln \sigma^\prime
 \\
 \beta_\theta &=& \frac{d\theta^\prime}{d\ln\lambda}= -g_1 (\sigma^\prime)
 W^N (\sigma^\prime) \sin \theta^\prime \label{pseudo-beta-theta}
\end{eqnarray}
where the $W(\sigma^\prime)$ function is the same as in Eq. \eqref
{W-w-strong} and
\be
 g_1 (\sigma^\prime) = 8 \left( \frac{\pi N}{12 \ln^2
 \sigma^\prime}\right)
 \left[ \frac{\pi N}{12 \ln^2 \sigma^\prime}
 - 1 \right],~~~~~~ e^{-\sqrt{12N/\pi}} \ll \sigma^\prime \ll 1 . \label{g-1}
\ee
\subsubsection{Relation to instanton phase \label{relation-beta-sigma}}
These expressions are of the same general form as the instanton
results of Eqs \eqref{weakinstanton1} and \eqref{weakinstanton2}
that are valid in the range $\sigma^\prime \gg 1$. It is easy to
see, for example, that the higher order terms in Eq.
\eqref{pseudo-beta-theta} are all of the same type as the weak
coupling results described in Eqs \eqref{all-instanton2} and
\eqref{beta-n}. Unlike Eq. \eqref{all-instanton1}, however, our
results for the observable $\sigma^\prime$ or the function
$\beta_\sigma$ do not display any dependence on $\theta^\prime$.
The reason being that in the definition of $\delta S$ in Eq.
\eqref{delta-S} we have neglected all the terms that couple the
$q$ matrix field and the $A_\mu$ vector field. It can be shown
that a more careful definition of the quantity $g^\prime$ or
$\sigma^\prime$ generally leads to the same geometrical series for
the $\beta_\sigma$ function as in Eqs \eqref{all-instanton1} and
\eqref{beta-n}. \cite{PruiskenBurmistrovShankar} In the regime of
interest the $\theta^\prime$ dependent terms in the $\beta_\sigma$
function are all insignificant relative to the leading order
result of Eq. \eqref{pseudo-beta-theta}, however, and they do not
alter the main conclusions of the present investigation.
%
%
\begin{table*}
\begin{tabular}{|c||c|c|c||c|}
 \hline  \hline
     &            \\
 $n$ & $\beta_\theta^{(n)} (0)$  \\
     &            \\
 \hline \hline
           &                       \\
  ~~$1$~~  &   ~~$-2.794$~~         \\
  ~~$2$~~  &   ~~$-0.633$~~           \\
  ~~$3$~~  &   ~~$-0.268$~~            \\
  ~~$4$~~  &   ~~$-0.146$~~            \\
  \vdots   & \vdots                \\
  ~~$n$~~  &   $-\frac{2}{\pi n^2}$ \\
           &                       \\
 \hline \hline
\end{tabular}
 \caption{Numerical values of $\beta_\theta^{(n)} (0)$, see text.} \label{table1}
\end{table*}
%
\subsubsection{Relation to quantum Hall phase}
Keeping these remarks in mind we next turn to the renormalization
group results obtained for the quantum Hall phase, Eqs
\eqref{beta-sigma} and \eqref{beta-theta}. We expand the latter in
a trigonometric series according to
\begin{eqnarray}
 \beta_\theta (\theta^\prime) &=& \frac{\theta^\prime}{\pi} \Bigl [ 2\pi -
 |\theta^\prime| \Bigr ]
 \ln \Bigl [ \frac{|\theta^\prime|}{2\pi - |\theta^\prime|} \Bigr
 ] = ~ \sum_{n=1}^\infty \beta_\theta^{(n)} (0) ~\sin n
 \theta^\prime . \label{beta-n-1}
\end{eqnarray}
The coefficients $\beta_\theta^{(n)} (0)$ are finite numbers
defined by
\be
 \beta_\theta^{(n)} (0) = \int_{-\pi}^{\pi} \frac{d\vartheta}{\pi} \sin(n\vartheta)
 \beta_\theta (\vartheta).
\ee
These results clearly show that the instanton functions
$\beta_\theta^{(n)} (\sigma^\prime)$ in Eq. \eqref{all-instanton2}
all have a well defined strong coupling limit $\sigma^\prime =0$.
In the same limit we find that all the coefficients
$\beta_\sigma^{(n)} (\sigma^\prime)$ in Eq. \eqref{all-instanton1}
are zero as expected.

In Table \ref{table1} we list the numerical values of the lowest
order coefficients $\beta_\theta^{(n)} (0)$. By truncating the
series in Eq. \eqref{beta-n-1} one generally retains the correct
fixed point structure of the theory. The exact exponent value $2$
in Eq. \eqref{exponent-2} is expanded in an infinite series
according to
\begin{equation}
 2 = \sum_{n=1}^{\infty} n (-1)^n \beta_\theta^{(n)} (0). \label{crit}
\end{equation}
By keeping only the lowest order term we obtain an approximate
exponent value of $2.8$ which in all respects is quite reasonable.
The renormalization behavior is therefore well represented if one
extends Eqs \eqref{pseudo-beta-sigma} and
\eqref{pseudo-beta-theta} to include the $n=1$ sector of the
quantum Hall phase according to
\be
 g_1 (\sigma^\prime) \approx \beta_\theta^{(1)} (0) = 2.794 \dots,  ~~~~~~~
 \sigma^\prime \ll e^{-\sqrt{12N/\pi}} . \label{g-1-a}
\ee
Notice that the results for the combination $g_1 (\sigma^\prime)
W^N (\sigma^\prime)$ in Eq. \eqref{pseudo-beta-theta} are quite
similar to those of the quantity $\phi_1 (\kappa) = v_1 (\kappa)
e^{-\kappa}$ of the free energy, Eq. \eqref{v-1}. Fig. \ref{FIG8}
illustrates the cross-over behavior of the present case as well.
\section{Summary and conclusion \label{Summary}}
%
\begin{table*}
\begin{tabular}{|c||c|c|c|}
\hline  \hline
 & & & \\
 &\emph{Quantum Hall phase}
 & \emph{Pseudo instanton phase}
 & \emph{Instanton phase}\\
 & & & \\
\hline \hline
 & & & \\
 $\sigma^\prime = \sigma^\prime (\lambda)$
 & $\displaystyle 0 \leq \sigma^\prime \ll e^{-\sqrt{12N/\pi}} $
 & $\displaystyle e^{-\sqrt{12N/\pi}} \ll \sigma^\prime \ll 1$
 & $\displaystyle 1 \ll \sigma^\prime $ \\
 & & & \\
\hline
 & & & \\
 $ W (\sigma^\prime )$
 & $\displaystyle e^{-\pi/(12 \ln^2 \sigma^\prime)}$
 & $\displaystyle e^{-\pi/(12 \ln^2 \sigma^\prime)}$
 & $\displaystyle 4 \pi e^{-\gamma-1/2}\sigma e^{-2\pi\sigma^\prime}$ \\
 & & & \\
\hline
 & & & \\
 $f (\sigma^\prime)$
 & $\displaystyle \sigma^\prime \ln \sigma^\prime $
 & $\displaystyle \sigma^\prime \ln \sigma^\prime$
 & $\displaystyle -\frac{1}{2\pi}$   \\
 & & & \\
\hline
 & & & \\
 $f_1 (\sigma^\prime)$
 & -
 & -
 & $\displaystyle {8(2\pi)^{1/2} e^{-1} N^{3/2}} ( \sigma^\prime )^2$      \\
 & & & \\
\hline
 & & & \\
 $g_1 (\sigma^\prime)$
 & $\displaystyle 2.794 \dots$
 & $\displaystyle 8 \left( \frac{\pi N}{12 \ln^2
 \sigma^\prime}\right) \left[ \frac{\pi N}{12 \ln^2 \sigma^\prime} - 1
 \right]$
 & $\displaystyle 8(2\pi)^{3/2} e^{-1} N^{5/2} (\sigma^{\prime})^2 $      \\
 & & & \\
\hline \hline
\end{tabular}
\caption{Generalized instanton results for the renormalization
group $\beta$ functions $\beta_\sigma =f (\sigma^\prime) -f_1
(\sigma^\prime) W^N (\sigma^\prime) \cos \theta^\prime $ and
$\beta_\theta= -g_1 (\sigma^\prime) W^N (\sigma^\prime) \sin
\theta^\prime$ describing the entire regime $0 \leq \sigma^\prime
< \infty$. An estimate for the function $f_1(\sigma^\prime)$ in
the pseudo instanton phase and the quantum Hall phase is beyond
the scope of the present investigation, see text.} \label{table2}
\end{table*}
%
In this investigation we have addressed the super universality
concept of the $\vartheta$ vacuum in the large $N$ expansion of
the $CP^{N-1}$ model. This has been done in three consecutive
steps. First, we study the general consequences of ``massless
chiral edge excitations" in the context of the grassmannian
$SU(M+N)/S(U(M) \times U(N))$ non linear sigma model. By
separating the fractional topological sectors from the integral
ones we obtain a general theory of massless edge excitations
together with the fundamental parameters (``physical observables"
or ``conductances") that define the renormalization behavior of
the $\vartheta$ vacuum. Secondly, we explicitly evaluate these
parameters employing a revised and adapted version of the usual
large $N$ steepest descend methodology. This leads to a
demonstration of the robust quantization of the Hall conductance
along with exact scaling results for the quantum Hall plateau
transitions. Thirdly, we employ the results recently obtained from
instantons in order to bridge the gap between the weak and strong
coupling sides of the large $N$ expansion. The most general way in
which the renormalization group $\beta$ functions can be expressed
is in terms of discrete topological sectors $n$ according to
\begin{eqnarray}
 \beta_\sigma (\sigma^\prime ,\theta^\prime) &=& \sum_{n=0}^{\infty} \beta_\sigma^{(n)}
 (\sigma^\prime) \cos n\theta^\prime \approx f (\sigma^\prime)
 - f_1 (\sigma^\prime) W^N (\sigma^\prime) \cos \theta^\prime \label{xx-all}\\
 \beta_\theta (\sigma^\prime ,\theta^\prime) &=& \sum_{n=1}^{\infty} \beta_\theta^{(n)}
 (\sigma^\prime) \sin n\theta^\prime \approx ~~~~~~~~
 - g_1 (\sigma^\prime) W^N (\sigma^\prime) \sin \theta^\prime
 \label{xy-all}
\end{eqnarray}
We have shown that by retaining only the lowest order terms in the
series the renormalization is well represented in the entire range
$0 \le \sigma^\prime < \infty$. In Table \ref{table2} we list the
generalized instanton functions $W(\sigma^\prime)$,
$f(\sigma^\prime)$, $f_1 (\sigma^\prime)$ and $g_1
(\sigma^\prime)$ obtained in the three different regimes that we
have considered. These results, together with the exact strong
coupling expressions of Eqs \eqref{beta-sigma} and
\eqref{beta-theta}, are the justification for the renormalization
group flow lines sketched in Fig. \ref{FIG4}.

It should be mentioned that our findings are in complete
accordance with the original work of Jevicki \cite{Jevicki} who
showed that the large $N$ steepest descend methodology and the
instanton methodology are formally the same. Our results
invalidate the historical ``large $N$ picture" of the $\vartheta$
vacuum, however, in particular the claims which say that
``discrete topological sectors" do not exist and ``instantons" are
irrelevant.\cite{Witten,Affleck80} These historical claims are
borne out of a mishandling of the massless chiral edge excitations
in the problem, as well as incorrect assumptions about the order
in which the thermodynamic limit and the large $N$ limit should in
general be taken.

Even though the idea of super universality of quantum Hall physics
has already been foreshadowed by the pioneering papers in the
field more than two decades ago, \cite{Pruisken84} over the years
this idea has nevertheless been confronted with many incorrect
expectations and conjectures
\cite{Affleck83,Affleck91,Verbaarschot,Weidenmueller,Zirnbauer}
that in one way or the other are all related to the historical
``large $N$ picture" of the $\vartheta$ angle. There is, for
example, the persistent belief in the literature which says that
the theory generally has no gapless excitations, not even at
$\vartheta=\pi$, and the quantum Hall effect somehow does not
truly exist. By demonstrating super universality as done in the
present investigation, we essentially establish a new paradigm for
our understanding of topological issues in quantum field theory in
general, and the experiment on the quantum Hall effect in
particular.
\section{Acknowledgements}
The author has benefitted from stimulating discussions with A.
Abanov, I.S. Burmistrov and A.M. Finkelstein. The research was
funded in part by the Dutch Science Foundations \textit{NWO} and
\textit{FOM} and the EU-Transnational Access program (RITA-CT
2003-506095).

\end{document}